\documentstyle[11pt,aaspp4,twoside]{article}

\lefthead{Brogan \& Troland}
\righthead{Magnetic Fields Toward W49}

\begin{document}

% LaTeX definitions.
\newcommand\HII{H\,{\sc ii}}
\newcommand\HI{H\,{\sc i}}
\newcommand\OI{[O\,{\sc i}] 63 $\mu$m}
\newcommand\CII{[C\,{\sc ii}] 158 $\mu$m}
\newcommand\CI{[C\,{\sc i}] 370 $\mu$m}        
\newcommand\SiII{[Si\,{\sc ii}] 35 $\mu$m}
\newcommand\Hi{H110$\alpha$}
\newcommand\He{He110$\alpha$}
\newcommand\Ca{C110$\alpha$}
\newcommand\kms{km~s$^{-1}$}
\newcommand\7{$\sim 7$~km~s$^{-1}$}
\newcommand\4{$\sim 4$~km~s$^{-1}$}
\newcommand\tf{25$\arcsec$}
\newcommand\fo{40$\arcsec$}
\newcommand\ft{15$\arcsec$}
\newcommand\cmt{cm$^{-2}$}
\newcommand\cc{cm$^{-3}$}
\newcommand\CeO{C$^{18}$O}
\newcommand\tCO{$^{12}$CO}
\newcommand\thCO{$^{13}$CO}
\newcommand\htco{H$_2$CO}
\newcommand\hto{H$_2$O}
\newcommand\Blos{$B_{los}$}
\newcommand\Bth{$B_{\theta}$}
\newcommand\Bm{$B_{m}$}
\newcommand\Bvm{$\mid\vec{B}\mid$}
\newcommand\BS{$B_S$}
\newcommand\Bscrit{$B_{S,crit}$}
\newcommand\Bw{$B_W$}
\newcommand\mum{$\mu$m}
\newcommand\muG{$\mu$G}
\newcommand\mjb{mJy~beam$^{-1}$}
\newcommand\jb{Jy~beam$^{-1}$}
\newcommand\thi{$\tau_{HI}$}
\newcommand\toh{$\tau_{OH67}$}
\newcommand\tmin{$\tau_{min}$}
\newcommand\tmax{$\tau_{max}$}
\newcommand\dv{$\Delta v_{FWHM}$}
\newcommand\va{$v_A$}
\newcommand\Np{$N_p$}
\newcommand\np{$n_p$}
\newcommand\gtsim{\raisebox{-.5ex}{$\;\stackrel{>}{\sim}\;$}}
\newcommand\We{$\mid{\cal W}\mid$}
\newcommand\Ms{${\cal M}_S$}
\newcommand\Mw{${\cal M}_w$}
\newcommand\Te{${\cal T}$}
\newcommand\Ps{${\cal P}_s$}
\newcommand\Ra{${\cal R}$}
\newcommand\Da{${\cal D}$}
\newcommand\xb{$x_b$}
\newcommand\xbt{$x_b^2$}

\title{VLA \HI\/ Zeeman Observations Toward the W49 Complex}

\author{C. L. Brogan\altaffilmark{1} and T. H. Troland}
\affil{University of Kentucky, Department of Physics \& Astronomy, 
Lexington, KY 40506-0055}
\altaffiltext{1}{Current address: National Radio Astronomy Observatory, P. O. Box O, Socorro, NM 87801}

\authoremail{cbrogan@aoc.nrao.edu} 

\begin{abstract}

We report VLA \HI\/ Zeeman observations toward the W49A star-forming region and
the SNR W49B.  Line of sight magnetic fields (\Blos\/) of 60 to 300 \muG\/ at
\tf\/ resolution were detected toward W49A at velocities of \4\/ and \7\/.  The
\Blos\/ values measured toward W49A show a significant {\em increase in field
strength with higher resolution} especially for the \4\/ \HI\/ component.  The
\HI\/ gas in the velocity range $-5$ to 25 \kms\/ toward W49A shows good agreement
both kinematically and spatially with molecular emission intrinsically associated
with W49A.  Based on comparisons with molecular data toward W49A, we suggest that
the \4\/ \HI\/ component is directly associated with the northern part of the
\HII\/ region ring, while the \7\/ \HI\/ component seems to originate in a lower
density halo surrounding W49A.  We estimate that the W49A North core is
significantly subvirial (2\Te\//\We\/$\sim 0.2$), and that the total kinetic +
magnetic energies amount to less than 1/3 of the total W49A North gravitational
energy.  These magnetic field results suggest that W49A North is unstable to
overall gravitational collapse in agreement with evidence that the halo is
collapsing onto the W49A North ring of \HII\/ regions.

The majority of the \HI\/ column density toward W49B comes from Sagittarius Arm
clouds along the line of sight at $\sim 40$ \kms\/ and $\sim 60$ \kms\/.  No
significant magnetic fields were detected toward W49B.  Comparison of the spectral
distribution of \HI\/ gas toward W49A and W49B suggests that evidence placing W49B
3 kpc closer to the sun (i.e.  at 8 kpc) than W49A is quite weak.  Although we
cannot place W49B at the same distance as W49A, we find the morphology of a $\sim
5$ \kms\/ \HI\/ component toward the southern edge of W49B suggestive of an
interaction.

\end{abstract}


\section{INTRODUCTION}

\begin{figure}[h!]
\centerline{
\epsfxsize=6.0in \epsfbox{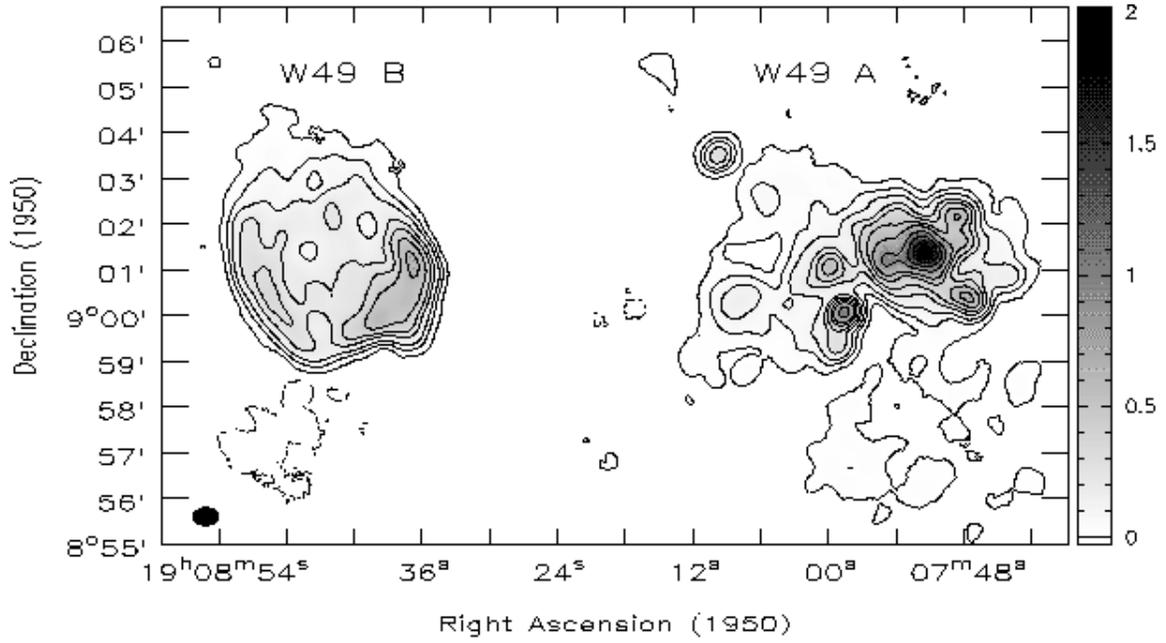}}
\figcaption[]{  Greyscale 21 cm continuum image of the W49 complex with the \HII\/ region
W49A to the west and the SNR W49B to the east.  Continuum contours at $-$15, 15, 50, 100, 200, 300, 450,
600, 900, 1200, and 1800 \mjb\/ are also shown.  The resolution is $26\farcs 5\times 23\farcs 9$ (P.A.
$-50.8\arcdeg$).\label{fig1}}
\end{figure}

The W49 complex is composed of one of the most luminous star forming regions in our galaxy (W49A) and a relatively
young supernova remnant (W49B) $12\arcmin$ to the east (see Fig.~\ref{fig1}).  At high resolution, W49A is
resolved into numerous compact \HII\/ regions (c.f.  Dreher et al.  1984; De Pree et al.  1997).  The \HII\/ regions
to the north (W49A North) form a ``ring'' of small ($\sim 5\arcsec$) ultra-compact \HII\/ regions which appear to
have formed during a period of triggered star formation (c.f.  Dickel \& Goss 1990; also see Fig.~\ref{fig2} for
a magnified view of W49 North).  The \HII\/ region with the highest emission measure {\bf G}\footnote{In this paper we use the
naming convention of De Pree et al.  (1997) for the W49A \HII\/ regions.}, is associated with a molecular outflow
and some of the strongest \hto\/ masers in the galaxy (Scoville et al.  1986; Walker, Matsakis \& Garcia-Barreto
1982).  To the SE of W49A North is W49A South which has a rim-brightened cometary shape and may have an edge-on
interface with dense molecular gas to the west (Dickel \& Goss 1990; also see Fig.~\ref{fig2}).  Gwinn, Moran, \&
Reid (1992) recently estimated the distance to W49A to be 11.4 kpc from \hto\/ maser proper motions.  The giant
molecular cloud associated with the W49A star forming region (as traced by \tCO\/) has a total mass of several
$10^5$ M$_{\odot}$ (Mufson \& Liszt 1977).

\placefigure{fig1}

Efforts to explain the complex morphology and kinematics of W49A have been abundant, and although progress has
been made, thus far no single model can reproduce its observed properties.  Low excitation molecular emission
lines (i.e.  \tCO\/($1-0$)) toward W49A are resolved into two components with velocities of \4\/ and 12 \kms\/
(Mufson \& Liszt 1977).  Such observations led to the suggestion that there are two distinct clouds in the
vicinity of W49A, and that their collision is responsible for triggering the observed star-formation (e.g.
Mufson \& Liszt 1977; Miyawaki, Hayashi, \& Hasegawa 1986; Serabyn, G\"usten, \& Schulz 1993).  However,
observations of high excitation (i.e.  \tCO\/($7-6$)) and low abundance (i.e.  \thCO\/) lines have revealed the
presence of colder, lower density gas at $\sim 7$ \kms\/ which appears as self-absorption in more abundant
species (Jaffe, Harris, \& Genzel 1987; Phillips et al.  1981).  Moreover, inverse P Cygni profiles with
absorption at $\sim 12$ \kms\/ have been observed in high resolution ($\sim 5\arcsec$) observations of HCO$^+$,
NH$_3$, and CS toward region {\bf G} which indicate the presence of infall (Welch et al.  1987; Jackson \&
Kraemer 1994; Dickel et al.  1999).  These facts have been taken to imply that the infalling gas at $\sim 12$
\kms\/ originates from a ``halo'' component at $\sim 7$ \kms\/ and is falling onto the the ring of \HII\/ regions
in W49A North.  Attempts to measure rotation in this gas or of the \HII\/ regions themselves have been largely
unsuccessful or unreliable (Welch et al.  1987; Jackson \& Kraemer 1994; De Pree, Mehringer, \& Goss 1997; Dickel
et al.  1999).  In addition to infall, there is also some evidence for the existence of more than one cloud 
or a velocity gradient in W49A (c.f. Dickel et al.  1999).
 
W49B is estimated to lie at a distance of $\sim 8$ kpc (e.g.  Moffett \& Reynolds 1994; also see \S 4.2) and has one
of the highest 1 GHz SNR surface brightnesses in the Galaxy.  It has a composite morphology with edge brightened radio
synchrotron emission but centrally concentrated thermal X-ray emission (c.f Moffett \& Reynolds 1994; Smith et al.
1985).  The spectral index of W49B is estimated to be $\alpha\sim -0.5$ ($S\propto \nu^{\alpha}$) and the X-ray
emission is thought to originate from a reverse shock in the remnant's interior (Green 1988; Kassim 1989; Moffett \&
Reynolds 1994).  The presence of such a reverse shock likely indicates that this remnant is young, with an age of
$\sim 6000$ years (Smith et al.  1985).  Moffett \& Reynolds (1994; and references therein) find that the synchrotron
emission of W49B has an unusually low fractional linear polarization, despite efforts to detect it at high frequencies
and resolutions.  These authors note that the low degree of linear polarization observed toward W49B may be indicative
of tangling in the magnetic field lines or Faraday depolarization within the SNR.

Diffuse atomic and molecular gas at LSR velocities of $\sim 40$ and 60 \kms\/ which does not appear to be physically
associated with W49 has also been observed toward W49A and W49B.  Some of the species observed in absorption include
\HI\/, OH, H$_2$CO, HCO$^+$ and CS (Lockhart \& Goss 1978, Bieging et al.  1982; Nyman 1984; Crutcher, Kaz\`es, \&
Troland 1987; Miyawaki, Hasegawa, \& Hayashi 1988; Greaves \& Williams 1994).  The excitation requirements of these
species suggest molecular hydrogen gas densities in the $10^2$ to $10^4$ cm$^{-3}$ range.  The $\sim 40$ and 60
\kms\/ components are thought to arise from Sagittarius spiral arm clouds.  Since this line of sight passes through
the Sagittarius arm twice, it is difficult to determine whether these clouds lie at their near or far distances
(e.g.  Dame et al.  1986; Jacq, Despois, \& Baudry 1988; also see \S 4.2).

Given the wide range of physical conditions that can be sampled by a single observation toward W49 -- the complex
kinematics of W49A, the SNR W49B, and diffuse gas at $\sim 40$ and 60 \kms\/ coupled with the limited resolution of
previous W49 \HI\/ observations ($2\arcmin$; Lockhart \& Goss 1978) we have performed VLA\footnote{The National
Radio Astronomy Observatory is a facility of the National Science Foundation operated under a cooperative agreement
by Associated Universities, Inc.}  \HI\/ Zeeman observations toward the W49 complex.  In particular, we hope to shed
some light on the physical processes governing the star formation process(es) in W49A by imaging the magnetic field
strengths and morphology in this intriguing source.  Previous VLA \HI\/ Zeeman magnetic field observations toward
star forming regions like W3 (Roberts et al.  1993) and M17 (Brogan et al.  1999) have revealed line of sight field
strengths of several hundred \muG\/ in gas intrinsic to these sources, as well as, complex field morphologies.

The results of our VLA \HI\/ Zeeman absorption observations toward the W49 complex are presented below and are
organized in the following order:  details of the data reduction and observing parameters are given in \S 2;
discussion of the 21 cm W49A and W49B continuum morphology can be found in \S 3.1, while the properties of the
\HI\/ absorption toward W49A and W49B are described in \S 3.2; our \HI\/ Zeeman magnetic field results are
presented in \S 3.3; and finally, the implications of these data are discussed in \S 4.

\section{OBSERVATIONS} 

The \HI\/ Zeeman data reported here consist of 12 hours of VLA B-config.  and 7 hours of VLA D-config.  data.  The
key parameters for these observations are given in Table~\ref{t1}.  We observed both senses of circular
polarization simultaneously.  Since Zeeman observations are very sensitive to small variations in the bandpass, a
front-end transfer switch was used to switch the sense of circular polarization passing through each telescope's IF
system for every other scan.  In addition, we observed each of the calibration sources at frequencies shifted
by $\pm$ 1.2 MHz from the \HI\/ rest frequency to avoid contamination from Galactic \HI\/ emission at velocities
near those of W49.

\placetable{t1}

The AIPS (Astronomical Image Processing System) package of the NRAO was used for the calibration, imaging, cleaning,
and calculation of optical depths.  The right (RCP) and left (LCP) circular polarization data were calibrated
separately and later combined during the imaging process to make Stokes I$=$(RCP + LCP)/2 and Stokes V$=$(RCP $-$
LCP)/2 data sets.  Bandpass correction was applied only to the Stokes I data sets since bandpass effects subtract
out to first order in the Stokes V data.  These \HI\/ data were also Hanning smoothed during imaging to improve
their signal to noise (S/N).  This means that while the channel separation is 0.64 \kms\/, the velocity resolution
is only $\sim 1.3$ \kms\/.

The AIPS task IMAGR was used to create ``cleaned'' \HI\/ line+continuum data sets at four different resolutions:
$5\arcsec$, $10\arcsec$, \ft\/, and \tf\/.  The \tf\/ resolution images were also convolved to a resolution of
\fo\/.  Note that uniform weighting of these data produces a beam of $\sim 5\arcsec$, while natural weighting
produces images with $\sim 25\arcsec$ resolution.  Data at different resolutions were needed for the wide range
of analysis presented here.  For example, the $5\arcsec$ data were used exclusively to create high resolution 21
cm continuum images, the $10\arcsec$ W49A \HI\/ line data are compared to H$_2$CO and \CeO\/ data with similar
resolutions, while \ft\/ is the highest resolution for which were able to make positive magnetic field detections
due to S/N constraints.  The \fo\/ data were used to search for magnetic field detections with the greatest
sensitivity.  The RMS noise characteristics and flux density to brightness temperature ($T_b$) conversion factors
for each resolution are summarized in Table~\ref{t2}.

\placetable{t2}

Separate line and continuum data sets were created by estimating and removing the continuum emission in the image
plane using IMLIN for W49A and W49B separately.  In this case continuum subtraction in the image plane turned out to
be the best option for two reasons:  (1) a greater number of line-free channels could be identified for W49A and
W49B individually; (2) given the strength of the \HI\/ absorption lines toward W49 (saturated at many positions),
better S/N was obtained for the line data when the line+continuum were cleaned together (i.e.  where the line is
strong very little cleaning is needed -- reducing cleaning errors from channel to channel).

\section{RESULTS}

\subsection{21 cm continuum of the W49 complex}

Figure~\ref{fig1} shows our VLA 21 cm continuum image of W49A (\HII\/ region complex) and W49B (SNR) at $\sim
25\arcsec$ resolution.  This image reveals the main continuum features of W49A and W49B including the
numerous individual \HII\/ regions present toward W49A and the ``blow out'' morphology of SNR W49B
toward the NE.  Inside the 15 \mjb\/ contour level on Fig.~\ref{fig1}, the flux density from W49A is $\sim 24$
Jy, while the contribution from W49B is $\sim 17$ Jy.  The rms noise in this $25\arcsec$ image is $\sim
3$ \mjb\/.  Evidently, about half of the 21 cm continuum flux is missing from our synthesis images since
single dish estimates of the flux densities of W49A and W49B at 21 cm are $\sim 45$ Jy and $\sim 25$ Jy,
respectively (Pastchenko \& Slysh 1973; Mezger et al.  1967).  Higher resolution continuum images of W49A
and W49B are discussed in detail below.

\subsubsection{W49A Continuum}

\begin{figure}[h!]
\centerline{
\epsfxsize=3.5in \epsfbox{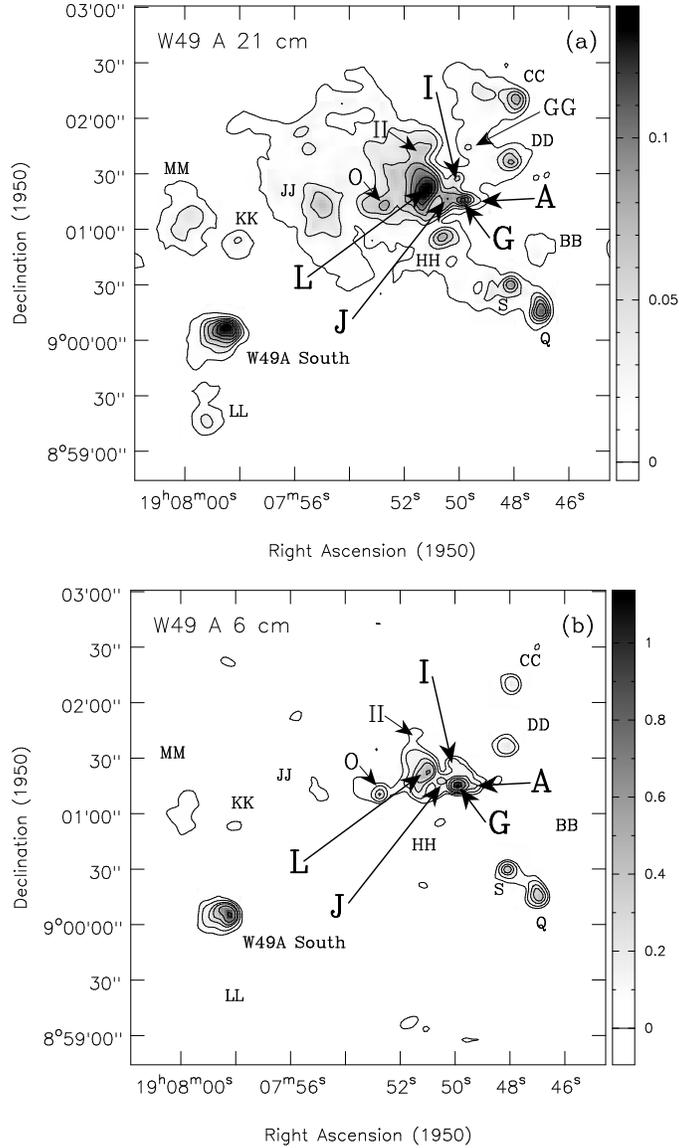}}
\figcaption[]{ W49 A continuum images with $5\arcsec$ resolution. (a) 
Greyscale 21 cm continuum image of W49 A with $5\arcsec$ resolution.  Continuum
contours are shown at 10, 30, 50, 70, 90, 110, and 130 mJy beam$^{-1}$.  (b) Greyscale 6 cm continuum
image of W49 A with $5\arcsec$ resolution.  Continuum contours are shown at 50, 100, 200, 300, 500,
and 900 mJy beam$^{-1}$ (original VLA data from Dickel \& Goss 1990).  The individual \HII\/ regions that
are bright at 21 cm are indicated using the naming convention of De Pree et al.  (1997) and are repeated
on the 6 cm image for reference.  The largest letter designations indicate the \HII\/ regions which make
up the ``ring''.  Notice that at 21 cm the continuum peaks toward region {\bf L}, while at 6 cm the continuum
peak is toward \HII\/ region {\bf G}.\label{fig2}} 
\end{figure}

Continuum images of W49A with $5\arcsec$ resolution at 21 cm and 6 cm are presented in Figures~\ref{fig2}a and b.
The 6 cm continuum data were obtained from the VLA archive and were originally reported in Dickel \& Goss (1990).
At this resolution, the wide array of distinct \HII\/ regions which make up the W49A complex become apparent (see
also Dickel \& Goss 1990; De Pree et al.  1997).  A number of these \HII\/ regions are indicated on
Figure~\ref{fig2} by their letter designations following the nomenclature of De Pree et al.  (1997).  Regions {\bf
A}-{\bf L} make up the ``ring'' of \HII\/ regions first described by Welch et al.  (1987) (see \S 1) and are
indicated by the largest letter symbols in Fig.~\ref{fig2} (due to space limitations only the regions prominent at
21 cm are labeled).  The \HII\/ regions that are most frequently mentioned by name in the text also have arrows
pointing toward them.  Also note that with the exception of regions W49A South, {\bf LL}, {\bf S}, and {\bf Q} the
region shown in Fig.~\ref{fig2} is commonly referred to as W49A North.

\placefigure{fig2}

The most noticeable difference between the 21 cm and 6 cm images presented in Fig.~\ref{fig2} is the lack of
extended structure present in the 6 cm image.  However, this deficit is simply the result of a lack of short
spacing information in the 6 cm VLA B-array data.  The most significant difference between the continuum emission
at these two wavelengths is that the 21 cm continuum emission peaks at \HII\/ region {\bf L}, as opposed to region
{\bf G} in the 6 cm image.  This apparent lack of agreement is due to the fact that most of the \HII\/ regions on
the western side of the ``ring'' are optically thick at 21 cm.  Indeed, Dickel \& Goss (1990) found that these
western \HII\/ regions are even somewhat optically thick at 6 cm from comparison with data at 2 cm (also see De
Pree et al.  1997).

\subsubsection{W49B Continuum}

\begin{figure}[h!]
\centerline{
\epsfxsize=3.5in \epsfbox{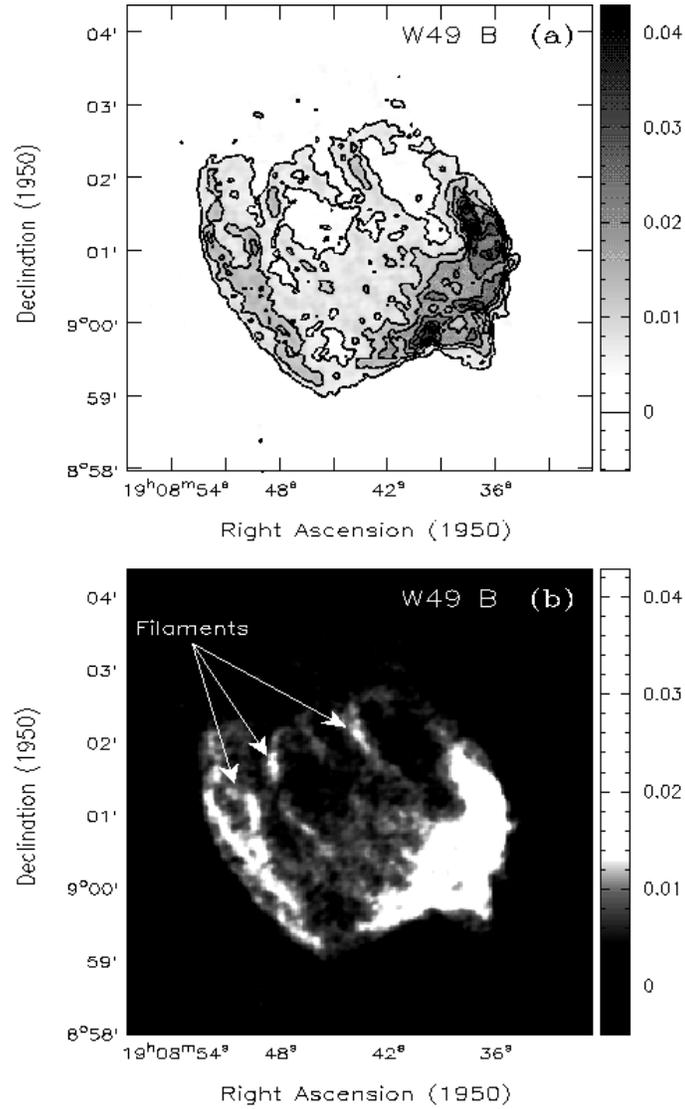}}
\figcaption[]{  Greyscale 21 cm continuum images of the SNR W49 B with $5\arcsec$
resolution.  (a) Continuum contour levels are 5, 10, 15, 20, 25, 30, 35, 40 \mjb\/.  (b) The greyscale
intensity has been adjusted to emphasize the continuum filaments visible in the northern and NE parts of the
SNR (see also Moffett \& Reynolds 1994).\label{fig3}}
\end{figure}

Figures~\ref{fig3}a and b show $5\arcsec$ resolution continuum images of W49B at 21 cm.  The continuum contours on
Fig.~\ref{fig3}a reveal the small scale synchrotron emission present in the SNR, while the greyscale of Fig.
3b emphasizes the low surface brightness filamentary and arc like structures present toward the eastern and
northern portions of W49B.  These filamentary features are also observed in the multiwavelength W49B continuum
images of Moffett \& Reynolds (1994).  Like these authors, we also note the arc-like appearance of the
central and western-most features along with the somewhat helical or twisted appearance of the eastern-most
filamentary structure (also see Fig.~\ref{fig6} in Moffett \& Reynolds 1994).  Moffett \& Reynolds (1994) rule out a
thermal origin for these filaments based on high resolution (15$\arcsec$) spectral index maps created from 90 cm,
21 cm, and 6 cm data.

\placefigure{fig3}

\subsection{\HI\/ Optical Depths}

A number of the \HI\/ components toward W49A (G43.2$-$0.0) and W49B(G43.3$-$0.2) suffer from severe saturation effects,
as might be expected from the low galactic latitude and distance to these sources ($\sim 8$ to 11 kpc).  Since
saturated channels have an undefined optical depth $[\tau = -$ln$(1+T_L/T_c)]$ (where $T_L$ is negative and $T_c$ is
positive; see also Roberts et al.  1995), saturation can cause significant underestimates of \HI\/ column
densities.  This is particularly true of W49A and W49B, because saturated pixels are present even at the continuum
peaks for some velocity components.  To mitigate this problem we have used
\begin{equation}
\tau_{lim}(\alpha,\delta)=-{\rm ln}\left[1 - \left(\frac{T_c(\alpha,\delta) - 3\sigma}{T_c(\alpha,\delta)}\right)\right]
\end{equation}
to estimate the largest optical depth value that can be reliably measured at each pixel given the continuum strength
($T_c$) and $\sigma$.  For the $15\arcsec$ (used for W49A) and $25\arcsec$ (used for W49B) data, $3\sigma\sim 10$
\mjb\/.  This ``limiting optical depth map'' was used to replace values in the original optical depth cube wherever the
calculated $\tau_{HI}$ is greater than the limiting value (i.e.  the calculated value is unrealistically large because
$\mid T_L\mid\sim T_c$) or the calculated value is undefined due to complete line saturation.  Note that the optical
depths in the channels replaced in this manner are {\em lower limits} to the true values.  In addition, the final
optical depth cubes were also masked wherever the continuum brightness is less than 50 and 30 \mjb\/ for the
$15\arcsec$ and $25\arcsec$ cubes respectively.

While the exact values of the optical depths derived from these ``corrected'' optical depth cubes retain a
degree of uncertainty (i.e.  they are lower limits), we feel that they better represent the distribution
of the \HI\/ gas.  However, two caveats should be kept in mind while examining the $N_{HI}/T_{s}$ images
presented in \S 3.2.  First, despite the obvious advantage of replacing the saturated optical
depths with a lower limit (instead of zero), this procedure can introduce a morphological bias which causes the
$N_{HI}/T_{s}$ images to artificially resemble the continuum if large regions of the source at a given
velocity have been replaced and also where the continuum is weak.  We have tried to minimize this bias by
identifying where this effect is present in the sections below.  Second, although we present estimates for the
\HI\/ spin temperature $T_{s}$ wherever possible (as deduced from previous observations), this is essentially
an unknown quantity, and need not be constant positionally within a single component or spectrally for
components of similar velocity (i.e.  the \HI\/ ``groups'' described below).  Therefore, $N_{HI}/T_{s}$ images
summed over multiple velocity components, in particular, may be biased toward components with low $T_{s}$.

\subsubsection{\HI\/ Optical Depths Toward W49A}

The molecular gas intrinsically associated with W49A lies in the velocity range $-5$ to 25 \kms\/ and is kinematically
complex (see eg.  Dickel et al.  1999; \S 1).  There appear to be 5 different heavily blended \HI\/ components in this
velocity range toward W49A, but no Gaussian fitting was attempted due to the complexity of the spectra.  The
approximate center velocities of these \HI\/ components are 4, 7, 10, 14, and 18 \kms\/ (see for example Fig.  5).
W49A \HI\/ optical depth channel images in the velocity range 0 to 21 \kms\/ with $15\arcsec$ resolution are shown in
Fig.~\ref{fig4}.  For comparison of the optical depth replacement scheme, the {\em top} set of panels in Fig.  4 show
$\tau_{HI}$ masked wherever the displayed channel was saturated or the calculated value exceeds the limiting value
(Eq.  1), but {\em without} any replacement, while the {\em bottom} panels have the same masking but saturated and
uncertain optical depths have been replaced with lower limits.  In both sets of panels, every third channel is
displayed.

Figure 4 shows that the gas near $\sim 4$ \kms\/ is concentrated toward the ring of \HII\/ regions and toward the SW
(regions {\bf S} and {\bf Q}).  In contrast, the $\sim 7$ \kms\/ gas is widely spread over all of W49A, with a peak
toward region {\bf J}.  The $\sim 10$ \kms\/ \HI\/ gas is almost as widely distributed, but shows a $\tau_{HI}$
minimum between the eastern and western sides of W49A.  The $\sim 14$ \kms\/ gas has a distribution similar to that at
10 \kms\/, but has an additional peak towards W49 South.  The lack of \HI\/ gas between the W49 North continuum peak
and the SE regions of W49A suggests that gas between 10 - 16 \kms\/ may arise from two spatially distinct clouds.  The
\HI\/ gas near $\sim 18$ \kms\/ has a morphology that is similar to that at $\sim 7$ \kms\/ (i.e.  widely distributed
over the whole source).  It is also noteworthy that while these velocity components have somewhat different overall
morphologies, all of them peak toward the western side of the W49A ring of \HII\/ regions.  The W49A optical depth
morphologies between $-5$ to 25 \kms\/ (Fig.~\ref{fig4}) show overall agreement with the $2\arcmin$ resolution \HI\/
optical depths images presented by Lockhart \& Goss (1978).

\placefigure{fig4}

Since these \HI\/ data were observed in absorption, they can in principle lie anywhere in front of the source along the
line of sight.  Therefore, comparison with molecular tracers in the velocity range $-5$ to 25 \kms\/ (which are
intrinsically associated with W49A) are needed to establish whether the \HI\/ gas in this velocity range is also
directly associated with this star-forming region.  Figure~\ref{fig5} shows a $10\arcsec$ resolution 21 cm continuum
image (dashed contours and greyscale) with solid integrated \CeO\/ ($2-1$) emission contours superposed (summed from
$-$5 to 20 \kms\/; Dickel et al.  1999).  Surrounding this image are $\tau_{HI}$, 6 cm ($10\times\tau_{H_2CO}$), and
($0.1\times$ \CeO\/) profiles toward several W49A \HII\/ regions.  The VLA 6 cm \htco\/ data presented here have
$10\arcsec$ resolution and were originally discussed in Dickel \& Goss (1990).  Similarly, the $12\arcsec$ resolution
IRAM \CeO\/ ($2-1$) data are also presented in Dickel et al.  (1999) and were obtained from the UIUC Astronomical
Digital Image Library.  

\placefigure{fig5}

Figure~\ref{fig6} shows the $10\arcsec$ resolution $N_{HI}/T_{s}$ integrated from $-20$ to 24 \kms\/ (with saturated
and uncertain optical depths replaced with lower limits as describe in \S 3.2).  The morphology of the $N_{HI}/T_{s}$
image shown in Fig.~\ref{fig6} is quite similar to the distribution of \CeO\/ ($2-1$) within the field of view
observed by Dickel et al.  (1999; see Fig.~\ref{fig5}) in that they both peak toward the western side of the ring of
\HII\/ regions.  The resemblance of $N_{HI}/T_{s}$ to the continuum toward the low surface brightness regions of W49A
is a result of the saturation replacement method.  The $\tau_{HI}$ profiles shown in Fig.~\ref{fig5} also have a
velocity extent that is quite similar to the molecular gas tracers H$_2$CO and \CeO\/ indicating that they most likely
arise from similar regions in W49A.  One exception is the \HI\/ gas near $\sim 18$ \kms\/ where there appears to be
little molecular gas toward any of the W49A \HII\/ regions (see Fig.~\ref{fig5}).  Therefore, excluding the $\sim 18$
\kms\/ gas, estimates of the spin temperatures ($T_{s}$) of the \HI\/ components between $-5$ to 17 \kms\/ can be
taken from previous dust and molecular line data of W49A.

\placefigure{fig6}

One reasonable estimate for the \HI\/ spin temperature ($T_{s}$) is the average CO($2-1$) kinetic temperature of $51\pm
5$ K estimated by Phillips et al.  (1981).  The $\sim 20\arcsec$ JCMT 450, 800, and 1100 \mum\/ images of W49A by
Buckley \& Ward-Thompson (1996) show that the dust emission peaks slightly north of region {\bf G}, with secondary
peaks toward the SW (regions {\bf S} and {\bf Q}), the NE (region {\bf JJ}), and W49A South.  The NE dust peak is
positionally coincident with the enhancement in $\tau_{HI}$ observed toward the eastern side of W49A between 10 and 16
\kms\/ (see Fig.~\ref{fig4}).  Buckley \& Ward-Thompson (1996) derive dust temperatures toward W49A North of 17 K, 50
K, and 145 K; the SE of 57 K and 210 K; and SW of 48 K and 240 K (also see Sievers et al.  1991).  These authors
suggest that the cold dust toward W49 North (17 K) originates from a halo surrounding W49A, while the warm dust at
$\sim 50$ K is associated with the compact \HII\/ regions.  The hot dust between $\sim 150$ to 250 K is thought to be
be a minor constituent of the total dust mass.  CO($7-6$) observations also indicate the presence of a warm gas
component ($80\lesssim T\lesssim 150$) toward W49A North extending $\sim 2\arcmin$, centered slightly north of \HII\/
region {\bf G} (Jaffe et al.  1987).  An upper limit to the \HI\/ $T_s$ can also be estimated from the lowest 21 cm
continuum brightness for which we can still observe \HI\/ absorption $\sim 150$ K (\tf\/), since the background
continuum source must be warmer than the foreground absorbing gas.  It seems likely then, that the \HI\/ spin
temperature in the $-5$ to 17 \kms\/ range is $20 \lesssim T_{s} \lesssim 150$ K.

Figures~\ref{fig7}a and b show $N_{HI}/T_{s}$ images for the velocity ranges 25.5 to 47.4 \kms\/ and 48 to 82 \kms\/.
The \HI\/ components in these two velocity ranges are thought to arise from Sagittarius Arm clouds (\S 1).  The
dominant \HI\/ velocity components in the $\sim 40$ \kms\/ group are at $\sim 34$ and $\sim 40$ \kms\/, while the $\sim
60$ \kms\/ group consists primarily of three components at $\sim 54$, $\sim 58$, and $\sim 64$ \kms\/.  Overall,
$N_{HI}/T_{s}$ for the $\sim 40$ \kms\/ and $\sim 60$ \kms\/ \HI\/ components is fairly uniform over W49A.  The
apparent concentration of somewhat higher $N_{HI}/T_{s}$ for the $\sim 40$ \kms\/ group toward the eastern regions of
W49A (W49A South, {\bf MM}, {\bf KK}) is real, as is the apparent increase of the $\sim 60$ \kms\/ $N_{HI}/T_{s}$
toward the NW.  However, these regions also suffered from the most saturation as indicated by the black boxes in Fig.
7, and the strong resemblance of these peaks to the W49A continuum morphology is due to the saturated optical depth
replacement scheme.

\placefigure{fig7}

It is clear from comparison of the greyscale ranges in Figs.~\ref{fig7}a, and b, that the $\sim 60$ \kms\/ group has the
greater overall $N_{HI}/T_{s}$.  This is because there are more \HI\/ components in the $\sim 60$ \kms\/ \HI\/
group (i.e.  greater velocity extent) since the individual \HI\/ lines in both groups have approximately equal
line depths.  The individual \HI\/ components in both the $\sim 40$ \kms\/ and $\sim 60$ \kms\/ velocity
ranges have linewidths of $\Delta v_{FWHM}\sim 5$ \kms\/.  As described above, a reasonable upper limit for $T_s$
can be obtained from the lowest continuum brightness for which we can still accurately measure $\tau_{HI}$.  In
the case of W49A, with \ft\/ resolution, this limit is 50 \mjb\/ or 150 K.

As mentioned in \S 1, gas at these velocities has been observed previously in numerous molecular species in
absorption against W49A.  A recent study of these Sagittarius spiral arm clouds using CS($2-1$) and CS($3-2$)
absorption by Greaves \& Williams (1994) with $\sim 20\arcsec$ resolution show that the velocities of the CS
components at $\sim 40$ \kms\/ and $\sim 60$ \kms\/ are quite similar to those observed in \HI\/, although the CS
line widths ($\sim 1$ to 2 \kms\/) are about three times narrower.  These authors suggest that the average densities
of these clouds are $10^2\lesssim n\lesssim 10^4$ \cc\/ for $10\lesssim T_{kin}\lesssim 50$ K based on the relative
intensities of the CS transitions.  Additionally, Greaves \& Williams find that the $\sim 40$ \kms\/ $N({\rm CS})$ is
higher than that of the $\sim 60$ \kms\/ CS gas, in contrast to the higher $N_{HI}/T_{s}$ found for the $\sim 60$
\kms\/ group in \HI\/.  The $\sim 40$ \kms\/ components are also measured to have the greater column density in
species like CN, NH$_3$, and SO; while the $\sim 60$ \kms\/ components have greater OH, \htco\/, and HCO$^+$ column
densities (Greaves \& Williams 1994; Tieftrunk et al.  1994; Nyman \& Millar 1989).  Evidently the column densities
measured for these clouds are sensitive functions of resolution, abundance, and/or density.

\subsubsection{\HI\/ Optical Depths Toward W49B}

The \HI\/ profiles toward W49B reveal the presence of four distinct ``groups'' of components at $\sim 60$, $\sim 40$,
$\sim 20$, and $\sim 5$ \kms\/.  Each group consists of several \HI\/ components that are heavily blended with each
other, but are clearly separated in velocity from the other groups.  Figures~\ref{fig8}a and b show \HI\/ optical
depth profiles toward the eastern and western 21 cm continuum peaks of W49B and Figures~\ref{fig9}a, b, c, d, and e
show $N_{HI}/T_{s}$ images for the total W49B \HI\/ velocity range and the $\sim 60$, $\sim 40$, $\sim 20$, and $\sim
5$ \kms\/ \HI\/ groups.  The \tf\/ resolution data were used for these analyses (and figures) to improve the S/N of
the data.  We note that the {\em unreplaced} column density images (see \S 3.2) in these velocity ranges have very
similar morphologies, but the magnitude of $N_{HI}/T_{s}$ at most pixels is significantly less, as expected if the
majority of the replaced channels lie near line centers.

The first thing that is obvious from these column density images is that the morphology of $N_{HI}/T_{s}$ does not
particularly follow the continuum, unlike the case for the gas associated with W49A (e.g.  Fig.  6).  Another striking
feature of these images is the ``spotty'' nature of the $N_{HI}/T_{s}$ morphology.  This is particularly noticeable in
the $\sim 20$ \kms\/ $N_{HI}/T_{s}$ image (Fig.  9d).  The cause of this ``spottiness'' is unkown, but could be
indicative of unresolved structures since the individual spots are about the size of the synthesized beam (\tf\/).  An
upper limit on their corresponding linear sizescale cannot be easily determined since the distance to each of the
\HI\/ groups is uncertain (ranging from $\sim 0.5$ to 11 kpc).  The total $N_{HI}/T_{s}$ image toward W49B
(Fig.~\ref{fig9}a) shows an increase in this ratio toward the western and SW sides of the remnant.  This SW region of
high \HI\/ column density is coincident with the region where Lacey et al.  ( 2000) observe free-free absorption of
the W49B 74 MHz continuum emission.  The kinematics, morphology, and degree of \HI\/ saturation for the individual
groups are described below.  We note that the {\em unreplaced} column density images in these velocity ranges have
very similar morphologies, but the magnitude of $N_{HI}/T_{s}$ at most pixels is significantly less, as expected if
the majority of the replaced channels lie near line centers.

\placefigure{fig8}

\placefigure{fig9}

The 60 \kms\/ group is composed of at least 5 components and is kinematically complex.  The $\sim 60$ \kms\/
$N_{HI}/T_{s}$ is shown in Fig.~\ref{fig9}b summed from 51.3 to 80.2 \kms\/.  With the exception of an \HI\/ component
at $\sim 74$ \kms\/ which is isolated to the NE, most of the \HI\/ gas in the $\sim 60$ \kms\/ group is concentrated
toward the western side of the remnant.  From comparison of Fig.~\ref{fig9}b with Fig.~\ref{fig9}a, it is clear that
the $\sim 60$ group dominates the total $N_{HI}/T_{s}$ toward W49B.  Downes \& Wilson (1974) observed an H134$\alpha$
radio recombination line toward W49B at a velocity of 65$\pm$10 \kms\/, indicating the presence of a partially ionized
medium or low surface brightness \HII\/ region at this velocity toward W49B.  Given the similarity of the highest
column density $\sim 60$ \kms\/ \HI\/ gas to the regions where low frequency free-free absorption was observed by
Lacey et al.  (2000), it seems likely that much of the \HI\/ gas in this group is intrinsically associated with the
absorbing medium.

The $\sim 40$ \kms\/ $N_{HI}/T_{s}$ image shown in Fig.~\ref{fig9}c is summed from 26.1 to 50.6 \kms\/.  The two main
components that make up the $\sim 40$ \kms\/ group are at $\sim 41$ \kms\/ and $\sim 34.5$ \kms\/.  The component
at $\sim 41$ \kms\/ is saturated toward the southern half of the remnant, while the $\sim 34.5$ \kms\/ component
is only saturated toward the northern part of the SNR.  Despite this saturation, it is clear that the $\sim 40$
$N_{HI}/T_{s}$ has an overall concentration toward the southern half of the remnant.  There is also a region of
enhanced $N_{HI}/T_{s}$ on the NW edge of the remnant which is due to a component at $\sim 45.5$ \kms\/ that is
only present toward this NW \HI\/ concentration (this component is not saturated).

The \HI\/ group at $\sim 20$ \kms\/ is comprised of two components and the $N_{HI}/T_{s}$ image shown in Fig.~\ref{fig9}d
is summed from 13.3 to 25.5 \kms\/.  The \HI\/ component at $\sim 21.0$ \kms\/ has a fairly uniform column
density and is not saturated.  The slight enhancement of this component to the SE does not appear to be real.  In
contrast, the other component in this group (at $\sim 16.5$ \kms\/) is saturated to the north but also shows a
significant column density enhancement to the north.

Perhaps the most intriguing \HI\/ group toward W49B lies at $\sim 5$ \kms\/ due to the coincidence of this velocity to
those found for the molecular gas intrinsically associated with W49A ($12\arcmin$ to the west).  The $\sim 5$ \kms\/
$N_{HI}/T_{s}$ image is shown in Fig.~\ref{fig9}e and is summed from $-4.8$ to 12.6 \kms\/.  The $\sim 5$ \kms\/
components are concentrated toward the extreme southern edge of the remnant.  This $\sim 5$ \kms\/ \HI\/ enhancement
is coincident with a steep gradient in continuum brightness which could be indicative of an interaction between the
SNR shock and the ambient medium.  The possible association of the $\sim 5$ \kms\/ W49B \HI\/ group with molecular gas
at similar velocities in W49A is unclear.  W49B is believed to lie some 3 kpc closer to the sun than W49A (e.g.
Radhakrishnan 1972), making an association between the \HII\/ region and the SNR unlikely.  However, this distance
estimate is somewhat uncertain and will be discussed further in \S 4.2.

\subsection{W49A \Blos\/ detections}

Stokes I and V cubes with \ft\/, \tf\/, and \fo\/ resolution were fitted for the line of sight magnetic field strengths
(\Blos\/) using the Zeeman equation (V$=2.8$\Blos\/d$I/$2d$\nu$ for \HI\/; see Troland \& Heiles 1982) and the MIRIAD
fitting program ZEEMAP.  The dependence of \Blos\/ on the strength of Stokes V and the derivative of Stokes I makes it
necessary to fit only one \HI\/ component at a time.  This can make it very difficult to fit heavily blended lines.
The complicated nature of the \HI\/ spectra toward W49A restricted our \Blos\/ search to those \HI\/ components that
are spectrally distinct (on at least one side) and therefore have unblended derivatives (on at least one side).  After
identifying which \HI\/ velocity components are good candidates in Stokes I, the Stokes V cube was searched in the same
velocity range for the distinctive Zeeman S shaped pattern (or half of it if the line is blended on one side) and then
fitted over the channel range that best defined that component.  To the ensure that the detections are real, each
\Blos\/ image is masked wherever the \tf\/ resolution 21 cm continuum image is less than 30 \mjb\/, and only those
detections with a signal to noise ratio \Blos\//$\sigma_{B_{los}} >$ 3 are considered significant.  Since random noise
spikes can cause spurious agreements between the derivative of Stokes I and Stokes V, we also require that the
sizescale of a given detection region equal or exceed the synthesized beam.  For these reasons, blanked or white
regions on the following \Blos\/ images only indicate regions where one (or all) of these conditions are not met, and
do not imply that \Blos\/ is zero or even small.

Significant line of sight magnetic fields toward W49A were detected in two different \HI\/ velocity components.  One of
these two \Blos\/ detection zones has a velocity of \4\/ and lies to the north of \HII\/ region {\bf G}, coincident
with regions {\bf I}, {\bf II}, and {\bf GG}, while the other at \7\/ covers much of the eastern 1/3 of W49A North and
is coincident with regions {\bf O} and {\bf JJ} (see Fig.~\ref{fig2} for \HII\/ region identification).  In this
context, `coincident' only implies that the \HII\/ regions are in the same direction as the the \Blos\/ detection
region.  A composite image showing both the \4\/ and \7\/ \Blos\/ with {\bf \fo\/ resolution} is shown in
Figure~\ref{fig10} and {\bf \tf\/ resolution} in Figure~\ref{fig11}.  Several features of these \Blos\/ images should
be noted:  (1) The detected \Blos\/ are in the range $\sim 60$ to $\sim 300$ \muG\/; (2) {\em The \4\/ \Blos\/ is
negative, while the \7\/ is positive} (negative \Blos\/ indicates that the line of sight field points toward the
observer); and (3) {\em The higher resolution \tf\/ \Blos\/ detections shown in Figure~\ref{fig11} are significantly
stronger than those detected with \fo\/ resolution} (Fig.~\ref{fig10}), especially for the \4\/ component.  This last
point can be recognized by noticing that the highest \fo\/ \Blos\/ contour in Fig.~\ref{fig10} is $|150|$ \muG\/ while
this is the lowest \tf\/ \Blos\/ contour shown in Fig.~\ref{fig11}.  Indeed, although an image of the \ft\/ \Blos\/
detections are not shown (significant fits at this resolution are quite patchy due to poorer S/N), there is a small
region of significant \4\/ \Blos\/ at this resolution and the field strengths appear to be about a factor of two higher
yet.  

At \fo\/ resolution, in the absence of significant line blending, \Blos\/ $\gtrsim 30$ \muG\/ should have been
detectable toward the continuum peak of W49A (assumes a linewidth of $\sim 6$ \kms\/).  However, this lower limit for
detecting \Blos\/ increases in proportion to the inverse of the continuum intensity (i.e.  at half the \fo\/ continuum
peak, \Blos\/ must be greater than $\sim 60$ \muG\/ for it to be detectable in these data).  Likewise, at higher
resolution the continuum intensity per beam is less (if the source is resolved) and therefore, \Blos\/ is harder to
detect.  It should also be reemphasized that line blending can significantly inhibit the detection of \Blos\/
independent of S/N considerations.

\placefigure{fig10}

\placefigure{fig11}

Sample \Blos\/ fits are shown in Figures.~\ref{fig12}a, b, and c for the \4\/ component at \fo\/, \tf\/, and \ft\/
resolution.  Similarly, sample \fo\/, \tf\/, and \ft\/ resolution \7\/ \Blos\/ fits are shown in Figs.~\ref{fig9}a,
b, and c for a position near the center of the \7\/ \Blos\/ detection region, while Figs.~\ref{fig14}a, and b show
\fo\/ and \tf\/ resolution \7\/ \Blos\/ fits for a position near the top of the \7\/ \Blos\/ detection region.  As
described above, the need to fit one velocity component at a time significantly restricts the number of channels
that can be fit for a particular component.  In all three figures, the channel range used in the fit is indicated by
the solid portions of the Stokes I and d$I/$d$\nu$ profiles.  Likewise, the position, resolution, and fitted value
of \Blos\/ $\pm\sigma_{B_{los}}$ is printed near the top each figure.  The increase of \Blos\/ with resolution
described above can also be seen in these sample \Blos\/ fits.  Unfortunately, the continuum flux and therefore
\HI\/ absorption toward the northern 7 \kms\/ position is too weak to detect \Blos\/ at \ft\/ resolution.

\placefigure{fig12}

\placefigure{fig13}

\placefigure{fig14}

\subsection{W49B \Blos\/ (non)detections}

Similar efforts were also made to detect \Blos\/ fields toward W49B, but were unsuccessful.  We estimate that in the
absence of significant line blending, \Blos\/ $\gtrsim 40$ \muG\/ should have been detectable in the \fo\/ resolution
\HI\/ data at the W49B continuum peak (assuming a linewidth of $\sim 5$ \kms\/).  Therefore, our lack of W49B \Blos\/
detections is in agreement with the single dish OH 1665 MHz Zeeman results of Crutcher et al.  (1987) who tentatively
detected a \Blos\/ of $\sim 20$ \muG\/ toward W49B at $\sim 60$ \kms\/ in OH absorption.

\section{DISCUSSION}

\subsection{Magnetic Fields in W49A}

\subsubsection{\Blos\/ at \7\/}

As described in \S 3.3, line of sight magnetic fields of $\sim 200$ \muG\/ are detected toward W49A North to the east
of region {\bf L} (see Fig.~\ref{fig2}) at an approximate velocity of \7\/ (also see Figs.~\ref{fig10} \&
~\ref{fig11}).  The \HI\/ component and (\Blos\/ detection) at \7\/ has a similar velocity to molecular emission line
peaks observed in the higher order transitions of \thCO\/, \CeO\/, and H$^{13}$CO$^+$ which are optically thin (e.g.
Mufson \& Liszt 1977; Jaffe et al.  1987; Nyman 1984).  This is different from the spectra of optically thick species
(i.e.  \tCO\/), which show instead double peaked profiles at \4\/ and $\sim 12$ \kms\/ (eg.  Miyawaki et al.  1986).
Jaffe et al.  (1987) suggest that the optically thin \7\/ molecular components originate from cool halo gas
surrounding W49A that is self-absorbed by the more abundant species like \tCO\/.  Indeed, their radiative transfer
modeling of CO($1-0$), CO($2-1$), and CO($7-6$) data require that $T\lesssim 50$ K and $n_{H_2}\lesssim 10^3$ \cc\/ in
order to explain the presence of \7\/ self-absorption in the CO($1-0$) and CO($2-1$) data along with its absence in
their $30\arcsec$ resolution CO($7-6$) data.

It seems quite plausible that the \HI\/ absorption observed at \7\/ could arise from such a halo, and it is certainly
the case that the $6-10$ \kms\/ \HI\/ optical depth channel images (Fig.~\ref{fig4}) show that \HI\/ gas is quite
widespread across W49A in this velocity range, as might be expected from a halo component.  From examination of the
Stokes I data cube, it also appears likely that the restriction of \7\/ \Blos\/ detections to the eastern parts of W49A
is a consequence of this component being more severely blended toward other parts of the source.  This point can be
appreciated by comparing the \7\/ component in Figs.~\ref{fig13}a, b, c to those in Figs.~\ref{fig12}a, b, c.  It is
clear from these figures that the \7\/ component is not very distinct toward the \4\/ detection region.

It is difficult, however, to reconcile the presence of magnetic field strengths up to $\sim 300$ \muG\/ in gas with
$n_{H_2}\lesssim 10^3$ \cc\/ if the currently widespread idea that the static and wave magnetic energies in
molecular clouds are approximately equal (eg.  Brogan et al.  1999; Crutcher 1999; Myers \& Goodman 1988a, b).  The
assumption of equipartition between the nonthermal, wave magnetic, and static magnetic energy densities implies that
\begin{equation}
B \approx 0.4 \Delta v_{NT} n_{p}^{1/2}~\mu {\rm G},
\end{equation}
where $\Delta v_{NT}\sim \Delta v_{FWHM}$ is the non-thermal linewidth in \kms\/, and $n_p$ is the proton number
density in \cc\/.  Assuming equipartition, $B\sim 300$ \muG\/, and $\Delta v_{NT}\sim 6$ \kms\/, Eq. 2
suggests that $n_{H_2}\sim 1\times 10^4$ \cc\/ -- in contrast to the $n_{H_2}\lesssim 10^3$ \cc\/ suggested by Jaffe
et al.  (1987).  This disagreement worsens further if the higher resolution magnetic field results showing even higher
field strengths (see Figs.~\ref{fig10} \& ~\ref{fig11}) are considered.  In addition, we are only sampling one
component of the field, so that $\mid{\bf\vec{B}}\mid$ could be even higher.

There are a number of ways that this apparent disagreement could be resolved in favor of equipartition.  For
example, if the temperature of the halo is reduced to $T\lesssim 25$ K, the upper limit on the density increases
to $n_{H_2}\lesssim 10^4$ \cc\/ (see Jaffe et al.  1987).  This result is in agreement with the suggestion by Dickel \&
Goss (1990) that the halo has a density of $\sim 5\times 10^4$ \cc\/ as inferred from 2 cm and 6 cm \htco\/
observations.  A lower temperature is also supported by the $\sim 20$ K dust temperatures estimated for the halo
by Buckley \& Ward-Thompson (1996) and Sievers et al.  (1991).  In addition, if the \HI\/ gas is somehow clumped
or compressed relative to most of the molecular gas it could originate in gas of higher density than the average
molecular density.  In fact, Dickel \& Goss (1990) see evidence for \htco\/ clumping on the sizescale of their
$\sim 2\arcsec$ beam toward W49A.  Conversely, equipartition may simply not exist in the halo component.

\subsubsection{\Blos\/ at \4\/}

Of the two \Blos\/ detections, the one at \4\/ appears the more closely related to the `ring' of \HII\/ regions in W49A
since it lies to the north of \HII\/ region {\bf G}, coincident with regions {\bf I}, {\bf II}, and {\bf GG}
(Figs.~\ref{fig2}, ~\ref{fig10}, \& ~\ref{fig11}).  There are a number of observations that indicate the presence of
molecular gas along the same line of sight as the \4\/ \Blos\/ detections.  For example, the \4\/ \Blos\/ detection
region is contained within the northern portions of the \CeO\/ ($2-1$) integrated intensity (see Fig.~\ref{fig15}) and
the \CeO\/ ($2-1$) profiles shown in Fig.~\ref{fig5} toward this region (i.e.  regions {\bf I} and {\bf II}) show
strong \CeO\/ emission at \4\/.  Additionally, the $\tau_{H_2CO}$ channel images of Martin-Pintado et al.  (1985) also
show a strong \htco\/ peak toward the \4\/ \Blos\/ region.  A similar peak is also observed in the \htco\/ data from
Dickel \& Goss (1990), and the CS ($2-1$) data of Dickel et al.  (1999).  Curiously however, the {\em velocity} at
which these species are strongest is between 8 \kms\/ and 15 \kms\/, {\em not \4\/} (see Region {\bf II} \htco\/
profile in Fig.~\ref{fig5}).  However, the \HI\/ line at 4 \kms\/ does not appear noticeably stronger than the other
\HI\/ components in this region either.

\placefigure{fig15}

Another clue about the properties of the \4\/ gas comes from the VLA radio recombination line velocities observed
toward the cospatial \HII\/ regions by De Pree et al.  (1997).  These authors measured H92$\alpha$ RRL velocities
of $\sim 9$ \kms\/ for regions {\bf I} and {\bf II}, while {\bf GG} has a H92$\alpha$ velocity of $\sim 15$
\kms\/.  This comparison suggests that the \htco\/ and CS($2-1$) gas is closely associated
with the \HII\/ regions, while the \4\/ \Blos\/ gas is blueshifted toward us.

One candidate to explain these kinematics is an outflow with a quiescent core region.  However, the only known
outflow in W49A North emanates from \HII\/ region {\bf G}, but in the opposite sense (i.e.  the northern lobe of
the outflow is {\em redshifted, not blueshifted}; see eg.  Scoville et al.  1986).  Additionally, Dickel \& Goss
(1990) suggest that the northern lobe of this outflow must be well behind the bulk of the molecular gas based on
\htco\/ absorption measurements.  Therefore, it seems that the outflow emanating from region {\bf G} is not
associated with the \4\/ \HI\/ gas.  However, given the relatively small velocity difference between the {\bf I},
{\bf II}, and {\bf GG} \HII\/ regions and the \4\/ \HI\/ gas (a separation of $\sim 5 - 11$ \kms\/) it is
entirely possible that a second outflow with its blue lobe pointing mostly in the plane of the sky (so that its
apparent Doppler shifted outflow velocity is small) has gone undetected.

From the similarity of 2 cm and 6 cm $\tau_{H_2CO}$ profiles toward W49A ($2\arcsec$ resolution), Dickel \& Goss (1990)
conclude that that the density of the absorbing molecular gas is fairly high with an average value of $n_{H_2}\sim
2\times 10^5$ \cc\/ (excluding the halo component at 6 to 10 \kms\/).  Using Eq.  2, $\mid{\bf\vec{B}}\mid\sim 400$
\muG\/ based on our \ft\/ resolution \4\/ Zeeman detections (see Fig.~\ref{fig12}), and a linewidth of 6 \kms\/, we
estimate that the average density is $n_{H_2}=1\times 10^4$ \cc\/.  This comparison may indicate that \Blos\/ would
rise even higher if it could be measured on smaller sizescales.  Since we have only measured the line of sight
component of the field, $\mid{\bf\vec{B}}\mid$ could be also be higher if the field is mostly in the plane of the sky.
Indeed, from Eq.  2, a proton density of $n_{p}\sim 4\times 10^5$ \cc\/ and linewidth of 6 \kms\/ implies that
$\mid{\bf\vec{B}}\mid\sim 1500$ \muG\/.  However, it is also possible that the \HI\/ gas is not well mixed with the
dense molecular gas so that using the H$_2$CO density is inappropriate.  In the future it may be possible to measure
the density and \Blos\/ of the gas simultaneously with molecular Zeeman tracers like CN, SO, or CCS using the
Greenbank telescope (GBT) or ALMA.

\subsubsection{Energetics of W49A}

With the new information provided by these magnetic field detections, together with information provided by
previous W49A observations, it is now possible to investigate the energy balance of W49A North.  To do this, it
is necessary to assume that the \HI\/ Zeeman detections reported here reflect the magnetic field strengths in the
bulk of the molecular gas in W49A.  This seems quite likely given the high field strengths measured toward W49A,
and the commensurately large gas densities implied by these data (see \S4.1.1 - 4.1.2). However, this assumption 
is fundamental to the following estimates, and should be kept in mind.

Buckley \& Ward-Thompson (1996) and Serabyn et al.  (1993) independently estimate from dust and CS observations
respectively, that there is approximately $1\times 10^5$ M$_{\odot}$ within the inner $1\arcmin$ (3.3 pc) of W49A
North.  Using these estimates, the gravitational potential energy of the inner $1\arcmin$ of W49A North (i.e.  R=1.7
pc) is \We\/$\sim 4\times 10^{50}$ erg.  By comparison, the thermal$+$turbulent kinetic energy in W49A North, assuming
a linewidth of 8 \kms\/, is only 2\Te\/$=3M\sigma^2\sim 7\times 10^{49}$ erg (where $\sigma$ is the one dimensional
velocity dispersion).  Therefore, since 2\Te\//\We\/ $\sim 0.2 < 1$, we find that the W49A North core is subvirial in
agreement with Miyawaki et al.  (1986) who found that the virial mass of the core is an order of magnitude smaller
than the observed mass.  If true, this implies that the W49A North core is subject to collapse unless there are other
means of support. We caution, however, that the parameter values estimated above are uncertain to a least factors of 
two and that these uncertainties could conspire to increase 2\Te\//\We\/. 

The idea that magnetic fields (both the static and wave components) can help support star-forming clouds against
overall collapse has long been suspected (see e.g.  McKee 1999).  In the case of W49A North we have detected average
line of sight magnetic field strengths (\Blos\/) of $\sim\mid 300\mid$ \muG\/ at \tf\/ resolution (even higher at
\ft\/ resolution) implying that a reasonable estimate for the total average field strength is
${\mid\bf\vec{B}}\mid\sim 600$ \muG\/.  The static magnetic energy resulting from a field of this magnitude is
\Ms\/=$(b/3)B_S^2R^3\sim 5\times 10^{48}$ erg (where $b$ depends on the field geometry and is $\sim 0.3$; see McKee
et al.  1999).  Moreover, since the wave component of the magnetic field is thought to be in equipartition with the
turbulent kinetic energy, there is an additional contribution to the magnetic energy of \Mw\/$\sim $\Te\/$\sim
3.5\times 10^{49}$ erg (see e.g.  Brogan et al.  1999; Zweibel \& McKee 1995).  From these estimates, it appears
that the the static and wave components of the total magnetic energy are {\em not} equal, in contrast to the results
for a number of other regions including the M17SW star-forming core (Brogan et al.  1999; see also Crutcher 1999).
Additionally, these estimates show that the total magnetic energy is only $\sim 0.1$\We\/ and that the magnetic and
kinetic energy together only add up to about a quarter of the gravitational energy.  Even if the total static field
strength is as high as $\mid{\bf\vec{B}}\mid\sim 1500$ \muG\/ as suggested in \S 4.1 based on estimates for the
density of the W49A core, \Ms\/$\sim 3\times 10^{49}$ erg and the magnetic plus kinetic energies still only add up to
$\sim 0.34$\We\/.

While it must be emphasized that the parameters used in the energy calculations above are fairly uncertain, the
gravitational energy in the W49A North core appears to be $\sim 3$ to 4 times larger than the combined magnetic plus
turbulent kinetic energies.  This imbalance suggests a natural explanation for both the copious amount of
star-formation on-going in W49A and the apparent overall collapse of the W49A North halo onto the ring of \HII\/
regions suggested by Dickel et al.  (1999).  Indeed, ignoring the external pressure, we estimate that the static
component of the magnetic field would need to be on the order of 5 mG to obtain \We\/ = 2\Te\/ + \Mw\/ + \Ms\/
(using \Mw\/$\sim $\Te\/).  While the increase in \Blos\/ with resolution observed in these data suggest that
tangling or small scale structure in the field could reduce the measured values of \Blos\/ (see e.g.
Figs.~\ref{fig10}, ~\ref{fig11}, ~\ref{fig12}, ~\ref{fig13}), and we are measuring only one component of the field,
it seems unlikely that these effects would amount to the necessary factor of 10 difference between the highest
observed \Blos\/ and ${\mid\bf\vec{B}}\mid$.

\subsubsection{Magnetic Field Morphology}

As shown in \S 3.3, the \4\/ and \7\/ \Blos\/ detections point in opposite directions (the \4\/ \Blos\/ points
toward us), and lie toward different regions of W49A (see Figs.  10 \& 11).  It is impossible to determine the
significance and implications of this morphology without knowing the plane-of-sky magnetic field direction and
magnitude, and the effect of line blending on detecting \Blos\/.  However, two facts are clear from the \HI\/ \Blos\/
data presented here:  (1) while \Blos\/ was only detected at \7\/ toward the eastern regions of W49A North, this is
also the region where the \7\/ \HI\/ gas is most distinct as a separate velocity component (see also \S 4.1.1); (2)
the \4\/ component is quite distinct in regions where a significant \Blos\/ at this velocity was NOT detected (compare
for example Figs.  12 \& 13).  From these facts, we suggest that the confinement of significant \7\/ \Blos\/ to the
eastern side of W49A may simply be the result of line blending further west, and that the strong \4\/ \Blos\/ fields
are restricted to the NW region of W49A.

In the absence of more data, particularly about the \7\/ \Blos\/ toward the same regions where the \4\/ \Blos\/ is
detected, there are many different scenarios that can explain the opposite signs of the \4\/ and \7\/ \Blos\/ in
different parts of the source.  For example, the negative direction of the \4\/ \Blos\/ could be the result of an
outflow like the one proposed in \S 4.1.2 to explain the kinematics of the \4\/ component (which is blueshifted with
respect to the nearby \HII\/ regions).  Alternantively, the infall of the halo gas reported by Dickel et al.  (1999;
also see \S 1) could also potentially play a role in the apparent field reversal.  Further progress in understanding
the magnetic field morphology of this complex source will have to await future linear polarization and molecular
Zeeman observations.

\subsection{Comparison of \HI\/ gas toward W49A and W49B}
 
Figure~\ref{fig16} compares the average \tf\/ resolution \HI\/ optical depth toward W49A with that observed toward
W49B.  Before calculating the average, the optical depth cubes were masked wherever the 21 cm continuum flux is less
than 30 \mjb\/.  The overall shapes of the average $\tau_{HI}$ profiles toward W49A and W49B shown in
Fig.~\ref{fig16} (upper panel) are very similar to those presented in Lockhart \& Goss (1978) using the Owens Valley
interferometer with $2\arcmin$ resolution.  However, the magnitudes of the average optical depth profiles shown here
are smaller by about a factor of 1.5.  This discrepancy is caused by our replacement of saturated and uncertain
optical depths with conservative lower limits.  This procedure underestimates $\tau_{HI}$, especially toward the
weaker parts of the continuum (see \S 3.2), causing an overall lowering of the average optical depth.  Since this
method was applied equally to W49A and W49B it should not change the relative differences between them.  Indeed, the
similarity of the average W49A($\tau_{HI})-$W49B($\tau_{HI}$) profile presented in Fig.~\ref{fig16} (lower panel)
to the one shown in Lockhart \& Goss (1978) suggests this is the case.

\placefigure{fig16}

Although the distance to W49A has been accurately determined to be 11.4 kpc from H$_2$O maser proper motion studies
(Gwinn et al.  1992), the distance to W49B remains uncertain.  Given that there is strong \HI\/ absorption up to
velocities of $\sim 75$ \kms\/ for both W49A (G43.2$-$0.0) and W49B (G43.3$-$0.2), it is clear that W49B must lie
beyond the tangent point at $\sim 6.2$ kpc (using galactic center distance of 8.5 kpc).  In addition, since the \HI\/
absorption toward W49B does not extend to negative velocities, the SNR must also lie within the Solar circle at $\sim
12.4$ kpc.  In the past, comparison of the average $\tau_{HI}$ profiles W49A and W49B have also been used to estimate
the distance to W49B.  The major difference between the average \HI\/ optical depths toward W49A and W49B is the
absence of \HI\/ absorption at velocities between $\sim 7$ to 14 \kms\/ and $\sim 50$ to 55 \kms\/ toward W49B (see
also Lockhart \& Goss 1978; Radhakrishnan et al.  1972).  Indeed, the greater overall optical depth toward W49A from $\sim 7$
to 14 \kms\/ and in the $\sim 60$ \kms\/ components has been used in the past to suggest that W49A lies farther away
than W49B by $\sim 3$ kpc (Kaz\`es \& Rieu 1970; Radhakrishnan et al.  1972).

We do not find clear evidence in the high resolution \HI\/ data present here that this is the case.  For example, the
$N_{HI}/T_s$ images in all velocity ranges toward both W49A and W49B show noticeable changes on size scales of $\sim
1\arcmin$ (see Figs.~\ref{fig6}, ~\ref{fig7}, and ~\ref{fig9}).  Therefore, given the $12\arcmin$ separation between
them, it is not surprising that there are significant variations in the average optical depth profiles toward W49A and
W49B regardless of their relative distances.  Indeed, although the average $\tau_{HI}$ from 50 to 58 \kms\/ is
stronger toward W49A, there appears to be more \HI\/ gas in the $\sim 58$ to 65 \kms\/ range toward W49B
(Fig.~\ref{fig16}).  Also, while there is a complete absence of \HI\/ absorption at velocities between $\sim 7$ to 14
\kms\/ toward W49B, there are a number of scenarios involving differences in kinematics, distribution, and temperature
that could explain this behavior without assuming that W49B is closer.  In any case, conversion of the excess W49A
$\sim 7$ to 14 \kms\/ $N_{HI}/T_s$ to the distance between W49B and W49A strongly depends on the assumed density and
temperature of the \HI\/ gas.  For example, if the $\sim 7$ to 14 \kms\/ \HI\/ absorption arises from the W49A halo
component as suggested in \S 4.1, ($n_{H_2}\sim 10^4$ \cc\/ and $T_s\sim 20$ K), then the necessary line of sight
separation between W49A and W49B reduces to $\sim 1$ pc compared to the 3 kpc obtained by Kaz\`es \& Rieu 1970 (also
see Radhakrishnan et al.  1972).  Summarizing, comparison of \tf\/ resolution \HI\/ absorption toward W49A and W49B
provides no clear evidence that W49B lies closer than W49A (i.e.  at 8 kpc).

\section{SUMMARY AND CONCLUSIONS}

\begin{figure}[h!]
\centerline{
\epsfxsize=5.5in \epsfbox{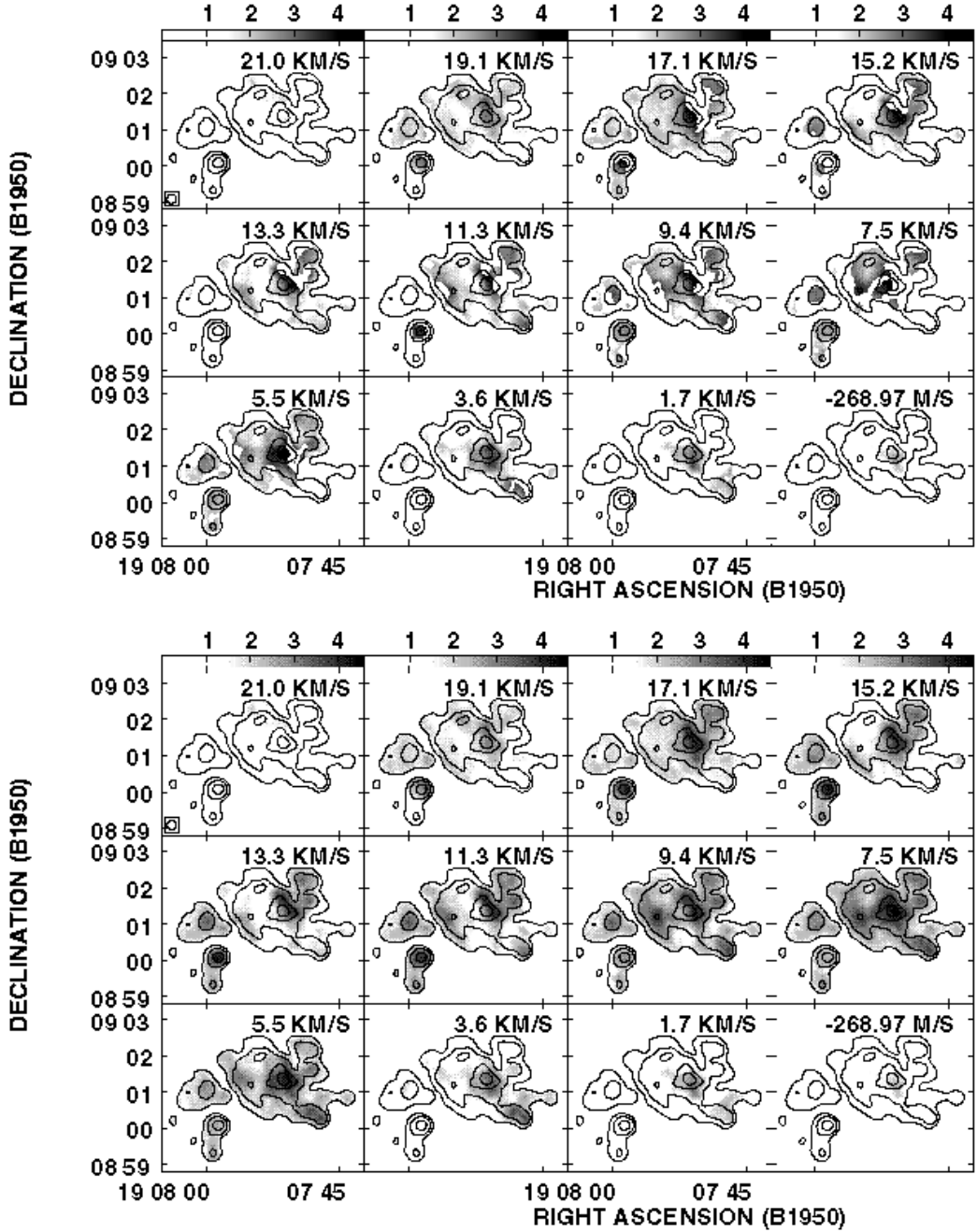}}
\figcaption[]{ Greyscale \HI\/ optical depth channel images toward W49A in the velocity range $-0.27$ to 21 \kms\/
(\HI\/ gas in this channel range appears to be directly associated with W49A).  The {\em top} set of panels show the
optical depths masked as described in \S 3.2 but {\em without} any replacement of saturated optical depths, while the
{\em bottom} panels have the same masking but saturated optical depths have been replaced with lower limits.  These
\HI\/ data have \ft\/ resolution and only every 3rd channel is shown.  The contours show the \ft\/ resolution 21 cm
continuum at 50, 150, 450, and 750 \mjb\/.\label{fig4}} \end{figure}

\begin{figure}[h!]
\centerline{
\epsfxsize=5.60in \epsfbox{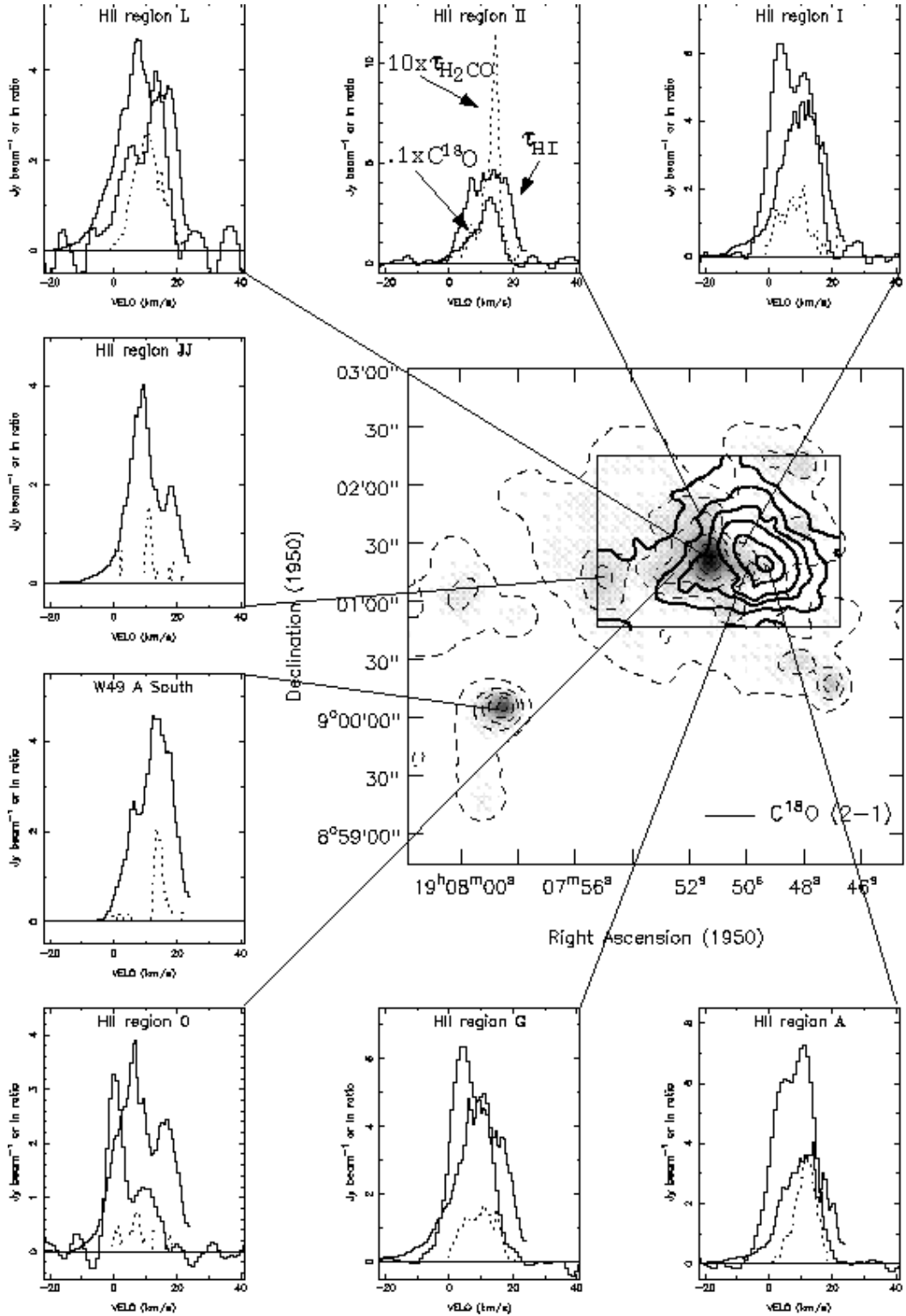}}
\figcaption[]{ Image of the W49A 21 cm continuum ({\em greyscale and dashed contours})
superposed with \CeO\/ ($2-1$) integrated emission contours summed from $-$5 to 20 \kms\/ ({\em solid contours}).
Surrounding this image are $\tau_{HI}$ ({\em thick black}), $10\times \tau_{H_2CO}$ ({\em dotted}) and
$0.1\times$ \CeO\/ emission ({\em thin black}) profiles toward several of the W49A \HII\/ regions.  The ({\em
dashed}) 21 cm contours are at 30, 130, 230, 330, and 430 \mjb\/.  The resolution of the 21 cm continuum,
$\tau_{HI}$ and $\tau_{H_2CO}$ data is $10\arcsec$, while the \CeO\/ data have a resolution of $12\arcsec$.
The VLA \htco\/ 6 cm data presented here were originally discussed in Dickel \& Goss (1990) and the IRAM
\CeO\/ data are also presented in Dickel et al.  (1999). Note that the \HI\/ optical depth profile 
for W49A South has a saturated (and replaced) channel at $\sim 17$ \kms\/.\label{fig5}}
\end{figure}

\begin{figure}[h!]
\centerline{
\epsfxsize=3.5in \epsfbox{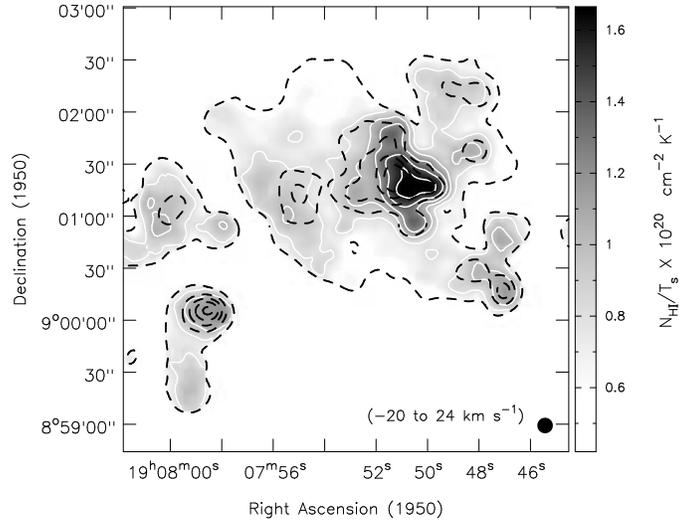}}
\figcaption[]{  Greyscale image of the lower limit of $N_{HI}/T_{s}$ toward W49A integrated from $-20$ to 24
\kms\/ with $10\arcsec$ resolution and white contours at (0.6, 0.8, 1.0, 1.2, 1.4, and 1.6) $\times 10^{20}$
cm$^{-2}$ K$^{-1}$.  21 cm continuum contours ({\em dashed}) at 30, 130, 230, 330, and 430 \mjb\/ with $10\arcsec$
resolution are also shown.\label{fig6}}
\end{figure}

The \HI\/ gas in the velocity range ($-5$ to 25 \kms\/) shows good agreement both kinematically and spatially with
molecular emission intrinsically associated with W49A (see Figs.~\ref{fig4} and ~\ref{fig5}).  Therefore, the
\HI\/ gas in this velocity range is likely to originate near the W49A star-forming complex.

Significant line of sight magnetic fields toward W49A were detected in two different \HI\/ velocity components.  Some
of the properties of these detections are:  (1) \Blos\/ of 60 to 300 \muG\/ were detected toward W49A at
velocities of \4\/ and \7\/; (2) {\em The \4\/ \Blos\/ is negative, while the \7\/ is positive} (negative \Blos\/
indicates that the line of sight field points toward the observer); and (3) The \Blos\/ values measured toward W49A
show a significant {\em increase in field strength with higher resolution} especially for the \4\/ \HI\/ component
(see Figs.~\ref{fig10}, ~\ref{fig11}, ~\ref{fig12}, ~\ref{fig13}, and ~\ref{fig14}).  Based on
comparisons of the \4\/ and \7\/ \Blos\/ detection regions and velocities with molecular data toward W49A, it seems
likely that the \4\/ \HI\/ component is directly associated with the northern part of the \HII\/ region ring, while
the \7\/ \HI\/ component seems to originate in a lower density halo surrounding W49A.

From previous molecular line and dust observations of W49A, we estimate that the W49A North core ($\sim 1\arcmin$ in
extent) is significantly subvirial, with 2\Te\//\We\/$\sim 0.2$.  In addition, it appears that the total magnetic
energy (assuming ${\mid\bf\vec{B}}\mid\sim 600$ \muG\/ and \Mw\/$\sim$ \Te\/) only contributes 0.1\We\/.  Adding the
different energy contributions, we estimate that the total kinetic + magnetic energies only amount to less than
about 1/3 of the total gravitational energy.  Indeed, we find that ${\mid\bf\vec{B}}\mid\sim 5$ mG is needed for
there to be overall equilibrium in W49A North (ignoring external pressure).  Therefore, while it must be emphasized
that the parameters used in these estimates are fairly uncertain, these new magnetic field results suggest that W49A
North is unstable to overall gravitational collapse as implied by evidence that the halo is collapsing onto the W49A
North ring of \HII\/ regions (e.g.  Dickel et al.  1999).

There are four main \HI\/ velocity groups detected toward W49B with velocities of $\sim 60$, $\sim 40$, $\sim 20$, and
$\sim 5$ \kms\/.  The majority of the \HI\/ column density toward W49B comes from the $\sim 60$ and $\sim 40$ \kms\/
\HI\/ components which arise from Sagittarius Arm clouds along the line of sight.  No significant \Blos\/ fields were
detected toward W49B.  Comparison of the spectral distribution of \HI\/ gas toward W49A and W49B (Fig.~\ref{fig16})
suggests that there is no clear evidence that W49B is 3 kpc closer to the sun than W49A (known to be at 11.4 kpc from
water maser proper motions; Gwinn et al.  1992).  Although we cannot place W49B at the same distance as W49A, we find
the morphology of the $\sim 5$ \kms\/ \HI\/ gas toward the southern edge of W49B suggestive of an interaction.
Indeed, gas at similar velocities is known to be intrinsically associated with W49A.

\acknowledgments

C.  Brogan would like to thank NASA/EPSCoR for fellowship support from the Kentucky Space Grant Consortium.  T.
Troland would like to gratefully acknowledge NSF support from AST-9988341.  We would also like to thank H.  Dickel,
W.  M.  Goss, C.  Lacey, and N.  Kassim for useful discussions on the nature of W49A and W49B and for providing
access to their data.  W.  M.  Goss also made many helpful comments and suggestions on the manuscript.

\newpage

\begin{figure}[h!]
\centerline{
\epsfxsize=3.5in \epsfbox{Fig7.epsxv}}
\figcaption[]{ Greyscale images of the lower limit of $N_{HI}/T_{s}$ toward W49A for (a) the $\sim 40$\kms\/ and (b)
the $\sim 60$ \kms\/ \HI\/ components toward W49A with \ft\/ resolution.  The $\sim 40$ \kms\/ $N_{HI}/T_{s}$ is summed
from 25.5 to 47.4 \kms\/ and the $\sim 60$ \kms\/ $N_{HI}/T_{s}$ is summed from 48 to 82 \kms\/.  The white
$N_{HI}/T_{s}$ contours are located at (0.55 and 0.65)$\times 10^{20}$ \cmt\/ K$^{-1}$ on (a) and (0.8, 0.9, and 1.0)
$\times 10^{20}$ \cmt\/ K$^{-1}$ on (b).  The black boxes superposed on (a) and (b) indicate regions where a
significant fraction of the integrated channels were saturated and therefore replaced with lower limits (see \S3.2).
21 cm continuum contours ({\em dashed}) with \ft\/ resolution are also shown at 50, 150, 450, and 750
\mjb\/.\label{fig7}}
\end{figure}

\begin{figure}[h!]
\centerline{
\epsfxsize=5.5in \epsfbox{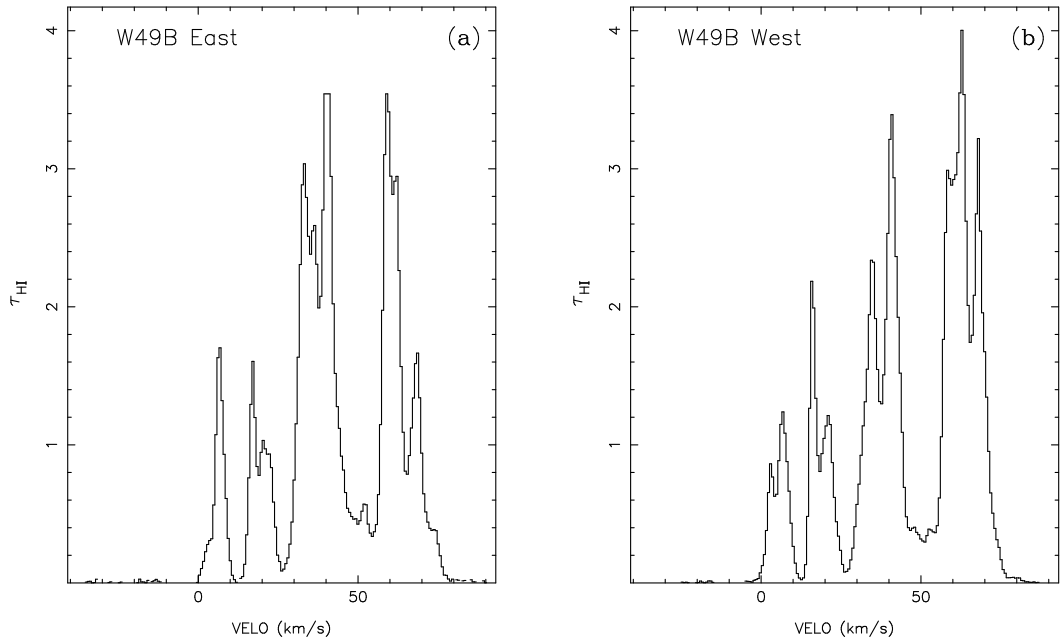}}
\figcaption[]{  \HI\/ optical depth profiles toward (a) the eastern W49B 21 cm continuum
peak and (b) the western W49B 21 cm continuum peak. These data have \tf\/ resolution and the separation 
between the two peaks is $\sim 3\arcmin$. The channel near $\sim 40$ \kms\/ in (b) is saturated and was replaced with 
a lower limit as described in \S 3.2.\label{fig8}} 
\end{figure}

\begin{figure}[h!]
\centerline{
\epsfxsize=6.0in \epsfbox{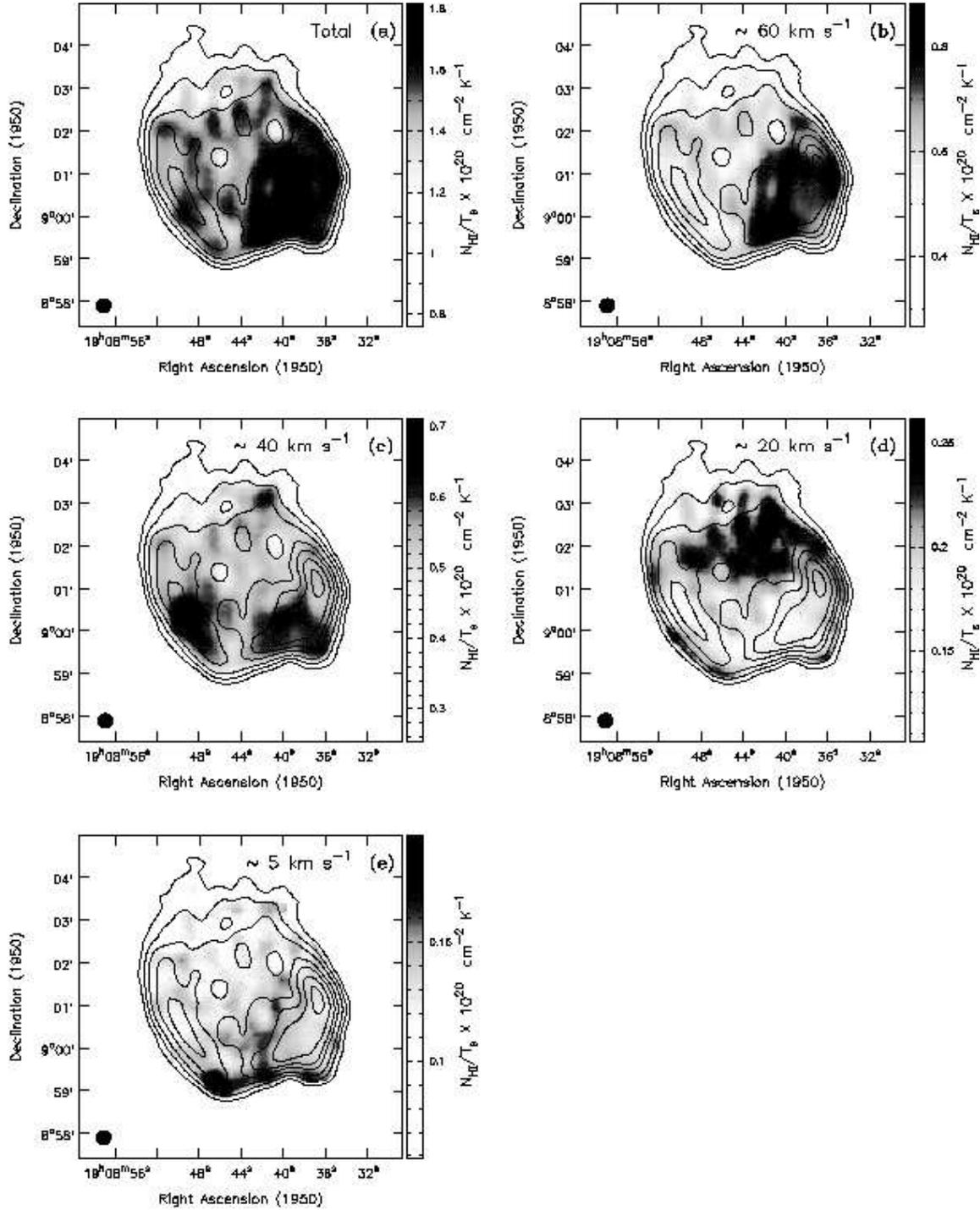}}
\figcaption[]{ Greyscale integrated images of the lower limit of $N_{HI}/T_{s}\times 10^{20}$ \cmt\/ K$^{-1}$ for the (a) total \HI\/
velocity range, (b) $\sim 60$ \kms\/ \HI\/ group, (c) $\sim 40$ \kms\/ \HI\/ group (d) $\sim 20$ \kms\/ \HI\/ group,
and (e) $\sim 5$ \kms\/ \HI\/ group toward W49B ($25\arcsec$) .  The black contours show the 21 cm continuum at
0.02, 0.05, 0.1, 0.2, 0.3, 0.45, and 0.6 \mjb\/ with \tf\/ resolution.\label{fig9}}
\end{figure}

\begin{figure}[h!]
\centerline{
\epsfxsize=5.0in \epsfbox{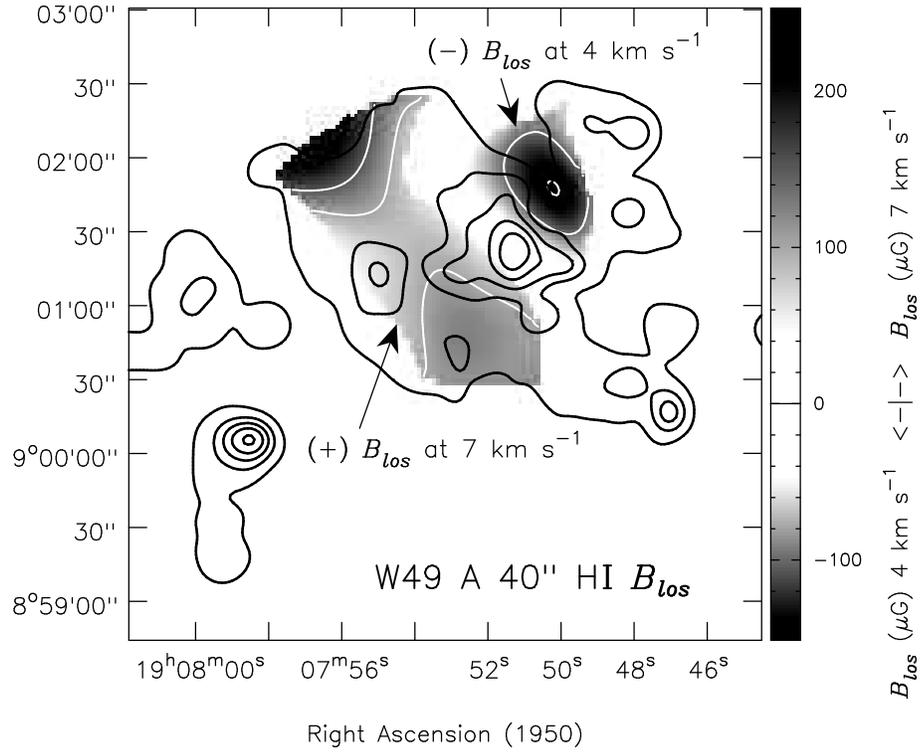}}
\figcaption[]{  Composite greyscale image of \Blos\/ for both the 4 \kms\/ and 7 \kms\/ \HI\/
components toward W49A with {\bf 40$\arcsec$ resolution}.  The black contours show the 21 cm continuum
at 40, 140, 240, 340, and 440 \mjb\/ with $10\arcsec$ resolution (same as Fig.~\ref{fig5}).  The white contours indicate
\Blos\/ at $-$100 and $-$150 \muG\/ for the 4 \kms\/ component and $+$100 and $+$150 \muG\/ for the 7
\kms\/ component.\label{fig10}} 
\end{figure}

\begin{figure}[h!]
\centerline{
\epsfxsize=5.0in \epsfbox{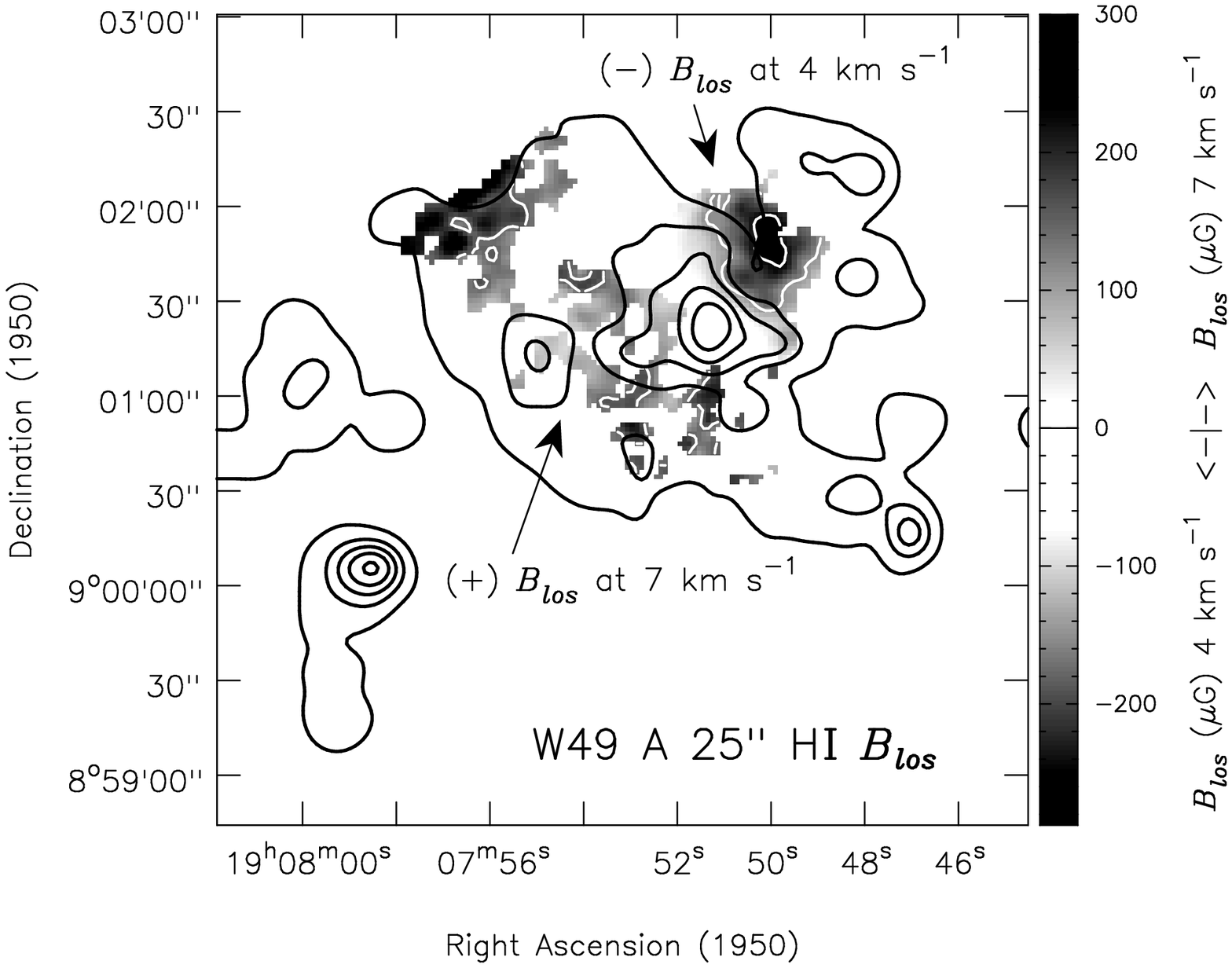}}
\figcaption[]{ Composite greyscale image of \Blos\/ for both the 4 \kms\/ and 7 \kms\/
\HI\/ components toward W49A with {\bf 25$\arcsec$ resolution}.  The black contours show the 21 cm continuum
at 40, 140, 240, 340, and 440 \mjb\/ with $10\arcsec$ resolution (same as Fig.~\ref{fig5}).  The white contours
indicate \Blos\/ at $-$150 and $-$250 \muG\/ for the 4 \kms\/ component and $+$150 \muG\/ for the 7 \kms\/
component.  The increase in \Blos\/ with resolution is particularly noticeable in the 4 \kms\/ component when
this image is compared to \Blos\/ in Fig.~\ref{fig10}.\label{fig11}}
\end{figure}

\begin{figure}[h!]
\centerline{
\epsfxsize=5.5in \epsfbox{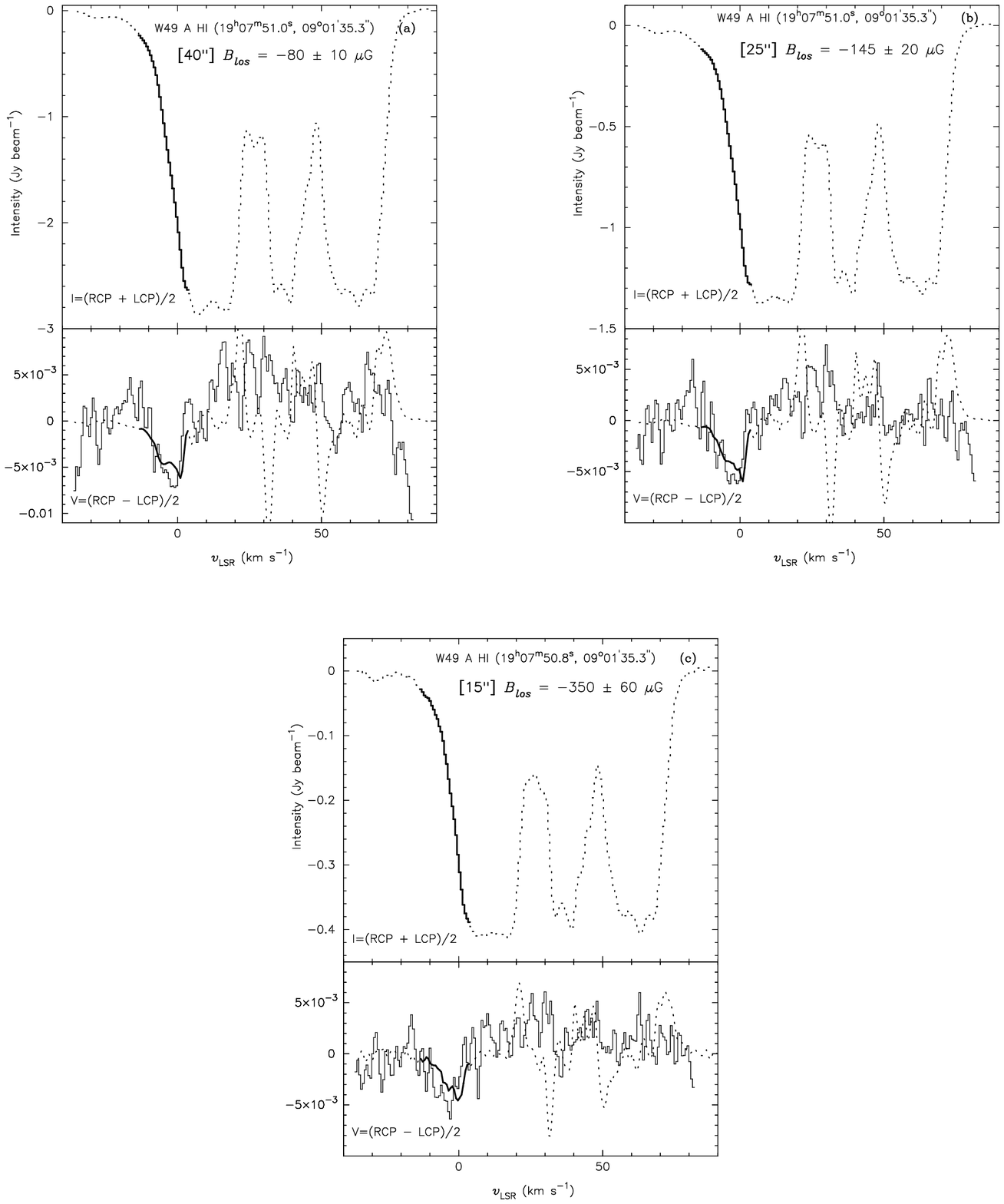}}
\figcaption[]{  \Blos\/ fits toward W49A at 4 \kms\/ (position is
$19{\rm^h}07{\rm^m}51.0{\rm^s}$, ${+09}\arcdeg$01\arcmin35\arcsec) with (a) \fo\/, (b) \tf\/, and (c)
\ft\/ resolution.  The upper panels show the VLA Stokes I profiles ({\em solid histogram}), and the
bottom panels show the VLA Stokes V profiles ({\em solid histogram}) with the fitted derivative of Stokes
I shown as smooth dotted curves.  The value of \Blos\/ fit for each position and its calculated error are
given at the top of each plot.  The solid portion of the Stokes I histogram ({\em upper panel})
shows the velocity range used in the fit.\label{fig12}} 
\end{figure}
%\newpage

\begin{figure}[h!]
\centerline{
\epsfxsize=6.0in \epsfbox{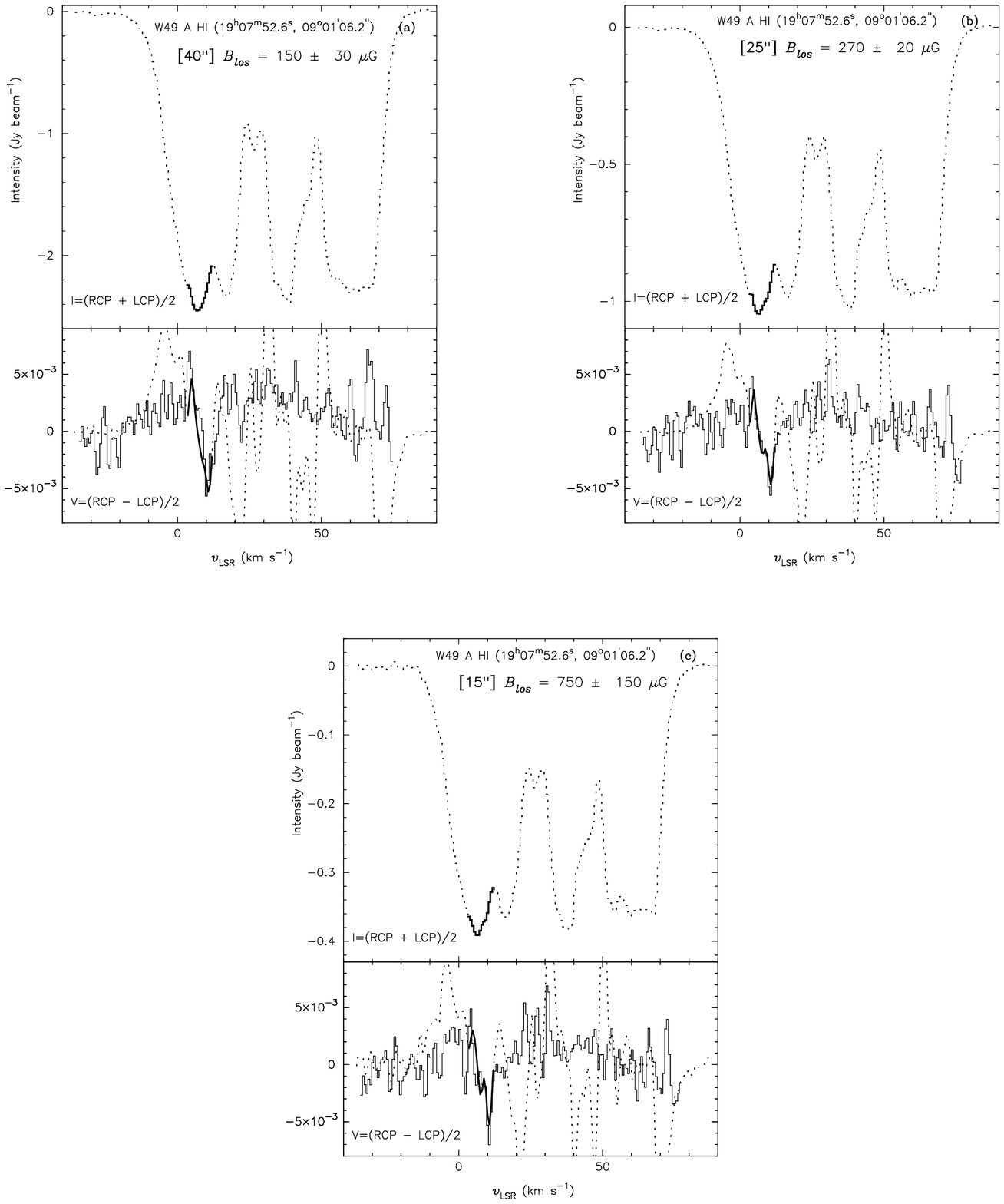}}
\figcaption[]{  \Blos\/ fits toward W49A at 7 \kms\/ for position 
$19{\rm^h}07{\rm^m}53{\rm^s}$, ${+09}\arcdeg$01\arcmin06\arcsec with (a)
\fo\/, (b) \tf\/, and (c) \ft\/ resolution.  The upper panels show the VLA Stokes I profiles ({\em solid
histogram}), and the bottom panels show the VLA Stokes V profiles ({\em solid histogram}) with the fitted
derivative of Stokes I shown as smooth dotted curves.  The value of \Blos\/ fit for each position and its
calculated error are given at the top of each plot.  The solid portion of the Stokes I histogram
({\em upper panel}) shows the velocity range used in the fit.\label{fig13}} 
\end{figure}

\begin{figure}[h!]
\centerline{
\epsfxsize=6.0in \epsfbox{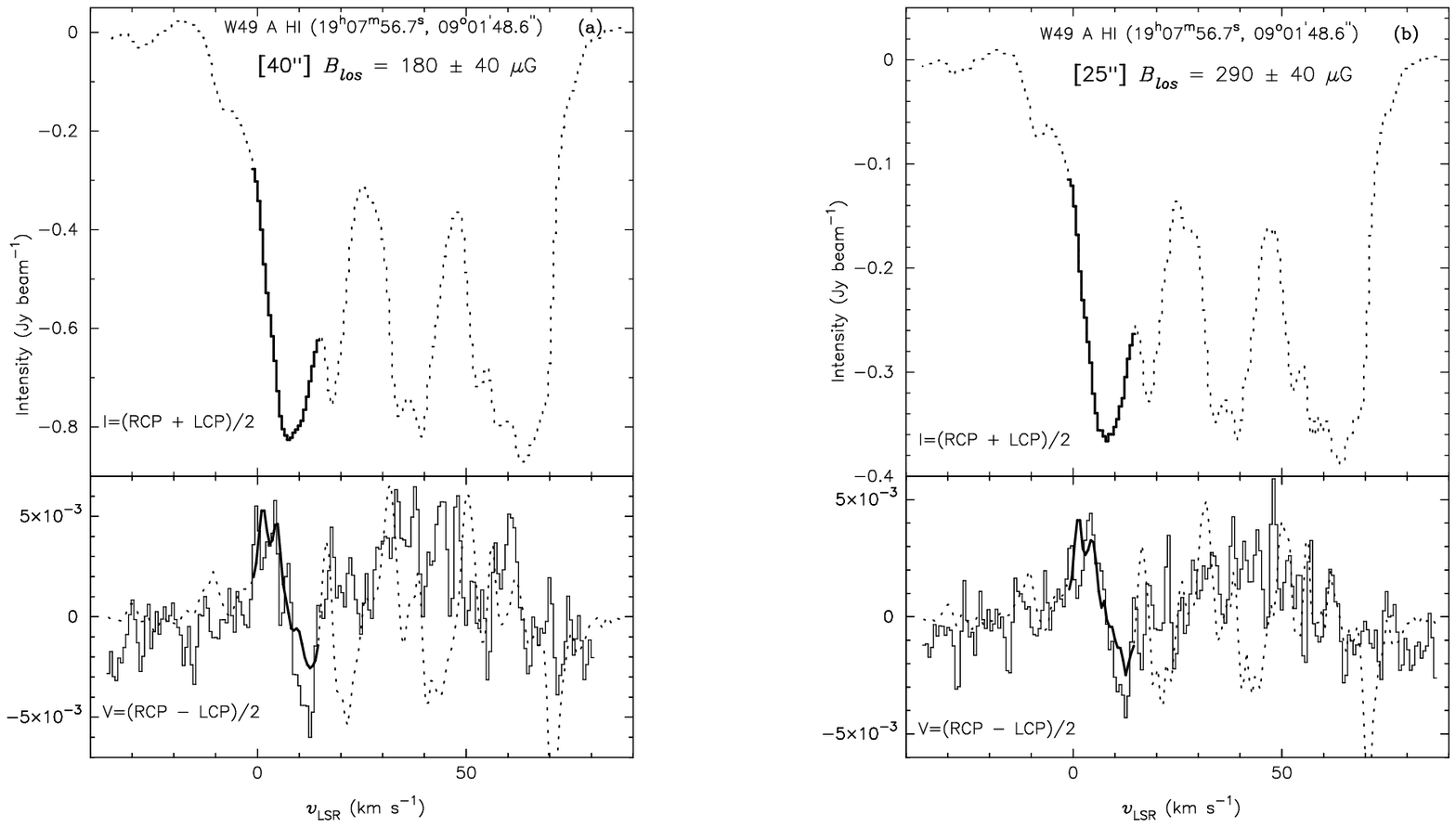}}
\figcaption[\Blos\/ fits toward W49A at 7 \kms\/ for position 
$19{\rm^h}07{\rm^m}57{\rm^s}$, ${+09}\arcdeg$01\arcmin49\arcsec.]{  \Blos\/ fits toward W49A at 7 \kms\/ for position 
$19{\rm^h}07{\rm^m}57{\rm^s}$, ${+09}\arcdeg$01\arcmin49\arcsec with (a)
\fo\/, and (b) \tf\/ resolution.  The upper panels show the VLA Stokes I profiles ({\em solid
histogram}), and the bottom panels show the VLA Stokes V profiles ({\em solid histogram}) with the fitted
derivative of Stokes I shown as smooth dotted curves.  The value of \Blos\/ fit for each position and its
calculated error are given at the top of each plot.  The solid portion of the Stokes I histogram
({\em upper panel}) shows the velocity range used in the fit.\label{fig14}} 
\end{figure}

\begin{figure}[h!]
\centerline{
\epsfxsize=5.0in \epsfbox{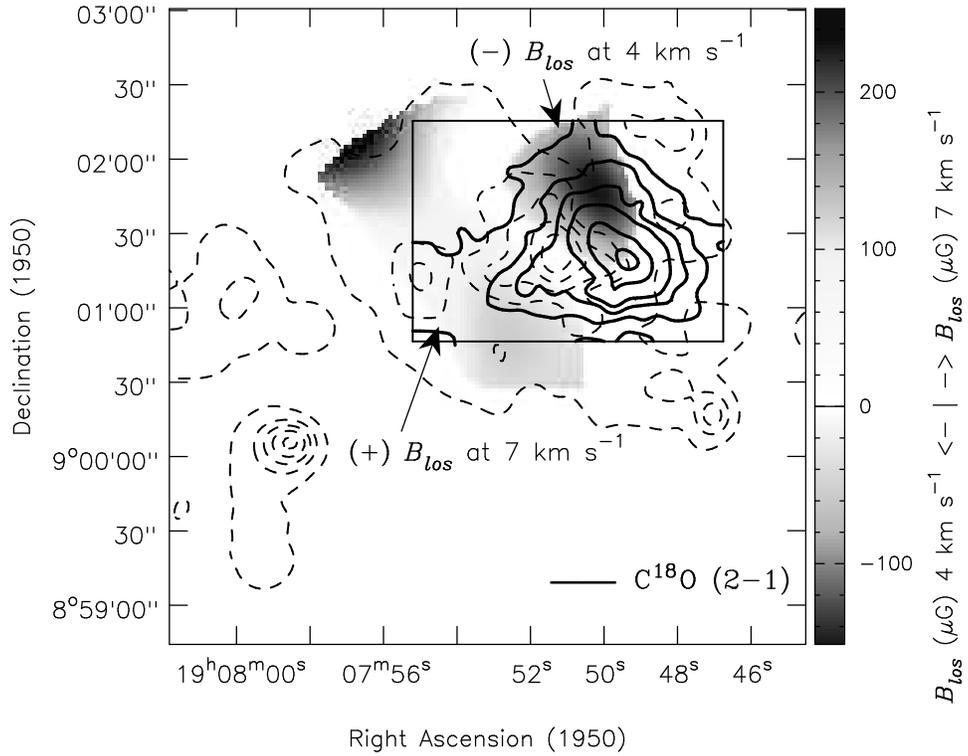}}
\figcaption[]{  Composite greyscale image of \Blos\/ for both the 4 \kms\/ and 7 \kms\/
\HI\/ components toward W49A with {\bf 40$\arcsec$ resolution}.  {\em Solid} black integrated \CeO\/ ($2-1$)
emission contours with $12\arcsec$ resolution (integrated from $-$5 to 20 \kms\/) are also shown (Dickel et
al.  1999; also see Fig.~\ref{fig5}).  The black ({\em dashed}) contours show the 21 cm continuum
($10\arcsec$ resolution) at 40, 140, 240, 340, and 440 \mjb\/ (same as Fig.~\ref{fig10}).\label{fig15}}
\end{figure}

\begin{figure}[h!]
\centerline{
\epsfxsize=3.5in \epsfbox{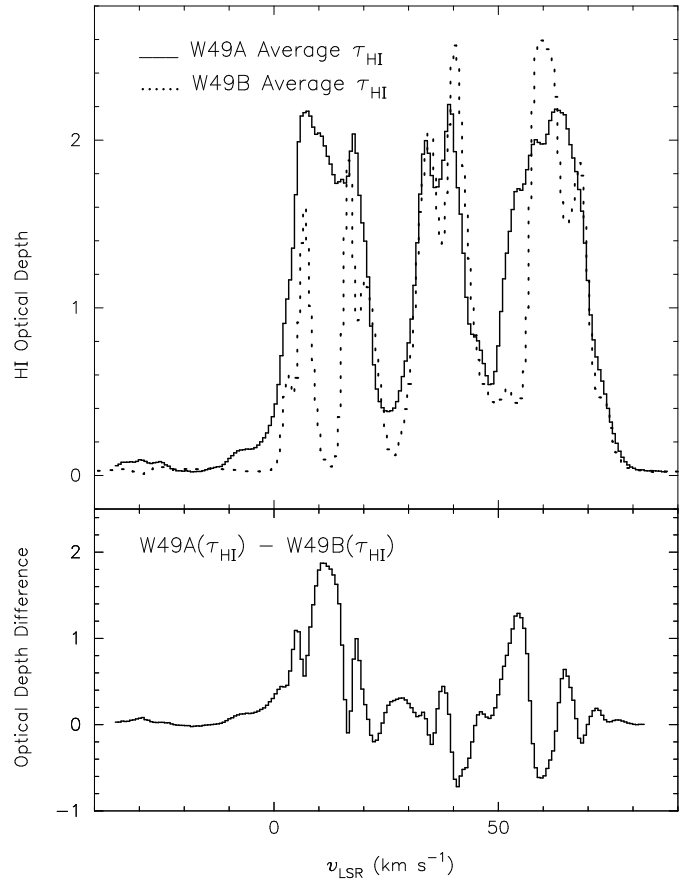}}
\figcaption[]{ {\em Upper panel}: Comparison of average \HI\/ optical depth profiles toward W49A ({\em solid}) and 
W49B ({\em dotted}). {\em Lower panel}: Difference between the average \HI\/ optical depth profiles 
shown in the upper panel [W49A($\tau_{HI})-$W49B($\tau_{HI})$].\label{fig16} } 
\end{figure}

\newpage

\begin{deluxetable}{lr}
\tablewidth{32pc}
\tablecaption{Parameters of VLA \HI\/ Zeeman Observations Toward W49}
\tablehead{
\colhead{Parameter}           & \colhead{Value}}      
\startdata
Frequency & 1420 MHz \nl
Observing date for D array data & Mar 12, 1999 \nl
Observing date for B array data & Jan 2, 2000 \nl
Total Observing time in D array & 7 hr \nl
Total Observing time in B array & 12 hr \nl 
Primary beam HPBW & 30\arcmin \nl
Phase and pointing center(B1950) & $19{\rm^h}08{\rm^m}20.0{\rm^s}$, 
${+09}\arcdeg$00\arcmin00\arcsec \nl
Frequency channels per polarization & 256 \nl
Total bandwidth & 781.25 kHz (165.05 \kms\/) \nl
Velocity coverage & -52.5 to +112.5 \kms\/ \nl
Channel separation & 3.052 kHz (0.64 \kms\/) \nl
Velocity resolution~$^a$ & 1.8 \kms\/ \nl
Angular to linear scale~$^b$ & $10\arcsec =$ 0.55 pc 
\label{t1}\enddata
\tablenotetext{a} {After Hanning smoothing.}
\tablenotetext{b} {Assumes distance of 11.4 kpc.}
\end{deluxetable}

\begin{deluxetable}{lccc}
\tablewidth{32pc}
\tablecaption{Parameters for W49 Images of Different Resolutions}
\tablehead{
\colhead{Resolution} & \colhead{RMS (line)} & \colhead{RMS (cont.)} & \colhead{$S_{\nu}\Rightarrow T_b~^a$}
\\ \colhead{($\arcsec$)} & \colhead{(mJy beam$^{-1}$)} & \colhead{(mJy beam$^{-1}$)} & \colhead{10~mJy~beam$^{-1}=T_b$(K)}}
\startdata
5 & \nodata & 8 & 272 \nl
10 & 5 & 5 & 68\nl
15 & 3 & 4 & 30 \nl
25~$^b$ & 2 & 3 & 11 \nl
40 & 4 & \nodata & 4 
\label{t2}\enddata
\tablenotetext{a} {Conversion factor to change 10 \mjb\/ into temperature in K for each resolution.}
\tablenotetext{b} {Exact resolution is $26\farcs 5\times 23\farcs 9$, P.A. $-50.8\arcdeg$. All other 
resolutions are exact with P.A. $=0.0\arcdeg$ after convolution.}
\end{deluxetable}

\end{document}